\documentclass[11pt,a4paper]{article} 
\usepackage{jheppub}
\usepackage[T1]{fontenc}

\usepackage{epsfig,multicol,bbm}
\usepackage{graphicx}
\usepackage{amsmath}
\usepackage{bm}
\usepackage{multirow}
\allowdisplaybreaks[1]

\newcommand\fverb{\setbox\fverbbox=\hbox\bgroup\verb}
\newcommand\fverbdo{\egroup\medskip\noindent%
			\fbox{\unhbox\fverbbox}\ }
\newcommand\fverbit{\egroup\item[\fbox{\unhbox\fverbbox}]}
\newbox\fverbbox

\def\beq{\begin{equation}}
\def\eeq{\end{equation}}
\def\ceq{\end{equation} \begin{equation}}
\def\bea{\begin{eqnarray}}
\def\eea{\end{eqnarray}}
\def\bei{\begin{itemize}}
\def\eei{\end{itemize}}
\def\bmat{\begin{matrix}}
\def\emat{\end{matrix}}
\def\ble{\begin{flushleft}}
\def\ele{\end{flushleft}}
\def\={\,=\,}
\def\+{\,+\,}
\def\-{\,-\,}

\def\GeV{\,{\rm GeV}\,}

\def\To{\Rightarrow}

\newcommand{\Fig}[1]{Fig.~\ref{#1}}
\newcommand{\Eq}[1]{eq.(\ref{#1})}

\title{Implications of LHC data on 125GeV Higgs-like boson for the Standard Model 
and its various extensions} 

\author[a]{Suyong Choi}
\affiliation[za]{Department of Physics, Korea University, Seoul 136-701, Korea}

\author[b]{Sunghoon Jung}
\affiliation[b]{School of Physics, Korea Institute for Advanced Study, Seoul 130-722, Korea}

\author[b]{P. Ko,}

\emailAdd{suyong@korea.ac.kr}
\emailAdd{nejsh21@kias.re.kr}
\emailAdd{pko@kias.re.kr}

\abstract{
Recent data on 125 GeV Higgs-like boson at the LHC starts to constrain the electroweak 
symmetry breaking (EWSB) sector of the SM and its various extensions.   
If one imposes the local gauge symmetry of the Standard Model (SM) ($SU(3)_c \times 
SU(2)_L \times U(1)_Y$) to the SM and any possible new physics  scenarios, 
the SM Higgs properties will be modified by intrinsically two different ways: 
by new physics either coupling directly to the SM Higgs boson $h$, or affecting indirectly
the SM Higgs properties  through the mixing of $h$ with a SM singlet scalar $s$. 
(Here $s$ is a singlet under the SM gauge group, but may be charged under a new 
gauge charge and can have nonrenormalizable couplings to non-SM particles.)
The models of two Higgs doublet, extra sequential and mirror fermions belong to the first category, whereas the models with a hidden sector dark matter, extra vector-like fermions 
and new charged vector bosons,  which can enhance the diphoton rate of the SM 
Higgs-like resonance, belong to the second category.  We perform a global fit to data 
in terms of the effective Lagrangian description of two interaction eigenstates  of scalar 
bosons, a SM Higgs and a singlet scalar, and their mixing. 
This framework is more suitable to study singlet-extended scenarios discussed above compared to other approaches based on the Lagrangian of mass eigenstates. With fairly model-independent assumptions, the effective Lagrangian contains at most four free parameters still encompassing the majority of models in the literature.  
Interestingly, the SM gives the best fit if all data from ATLAS and CMS are used,  
whereas various singlet extensions can fit better to individual ATLAS or CMS data.
Without further assumptions, an upper bound on the total width (or, non-standard 
branching ratio) is generically obtained. Furthermore, global fit based on our 
parameterization can be used to probe interactions of the singlet scalar if the singlet resides below $2m_W$. 
}

\begin{document} 
\maketitle
\section{Introduction\label{sec:intro}}

After the discovery of a new boson of mass around 125 GeV at the LHC
~\cite{Aad:2012tfa,Chatrchyan:2012ufa}, 
the most prompt question regarding this new particle is to identify its nature in particle 
physics context,  namely to verify if it is the SM Higgs boson that have been long sought for, 
or something else.  
For this purpose, its spin and parity~\cite{Choi:2002jk,Gao:2010qx,Ellis:2012wg,Englert:2012ct,Coleppa:2012eh,Bolognesi:2012mm,Choi:2012yg,Geng:2012hy,Ellis:2012jv,Djouadi:2013yb} and its couplings  to the SM particles should be determined as accurately as 
possible (see Ref.~\cite{Djouadi:2005gi} for the review on the SM Higgs boson and 
references therein). 

The recent results from ATLAS and CMS already tell us that the $J^{P}$ quantum 
number of 125 GeV boson is consistent with the SM prediction, namely 
$J^{P} = 0^{+}$ ~\cite{Chatrchyan:2012jja,Aad:2013xqa}.  
Other assignments of $J^{P}$ quantum number such as $J^P = 0^- $ or $2^+ , 2^-$ 
yield worse $\chi^2$ fits to the data compared to the $J^P = 0^+$ assignment, and 
thus disfavored.  

Determining the couplings of the 125 GeV resonance to the SM fermions and weak gauge 
bosons ($W^\pm , Z^0$) as well as gluons and photons has been studied by many groups 
\cite{Carmi:2012yp,Espinosa:2012ir,Giardino:2012ww,Ellis:2012rx,Espinosa:2012vu,Dawson:2012mk,Dawson:2012di,Low:2012rj,Giardino:2012dp,Ellis:2012hz,Espinosa:2012im,Carmi:2012in,Bonnet:2012nm,Bertolini:2012gu,Plehn:2012iz,Djouadi:2012rh,Batell:2012ca,Dobrescu:2012td,Cacciapaglia:2012wb,Corbett:2012ja,Belanger:2012gc,Cheung:2013bn,Cheung:2013kla,Ellis:2013ywa,Falkowski:2013dza,Giardino:2013bma,Contino:2013kra,Ellis:2013lra,Gainer:2013rxa,Belanger:2013xza,Joglekar:2012vc,Joglekar:2013zya,Belanger:2013kya,Maru:2013ooa,Maru:2013bja}
One usually considers a general case where the Higgs properties may be modified by 
new particles which can be described by dim-5 and dim-6 operators in the effective 
Lagrangian approaches~\cite{Manohar:2006gz}. 
In most of these approaches, it is assumed that all the new particles are heavy enough 
and can be integrated out, resulting in the higher dimensional operators and/or radiative 
corrections to renormalizable SM interactions.  This assumption works for the inert scalar doublet model,  the fourth generation sequential  or mirror fermions, 
or new colored and/or charged scalar fields.

However, in the presence of EW-scale singlet scalar boson that can mix with the SM Higgs 
boson, it may not be a good approximation to integrate out the singlet scalar bosons.
For example, in Refs.~\cite{Baek:2011aa,Baek:2012uj,Baek:2012se}, the authors demonstrated that
the effective Lagrangian approaches based on the unbroken subgroup ($SU(3)_C \times U(1)_Y$) of the SM gauge group produce erroneous results, especially on the direct 
detection cross section through the Higgs couplings to the dark matter particles. 
In the renormalizable versions of hidden sector (or Higgs-portal) DM models with the 
full SM gauge group ($SU(3)_c \times SU(2)_L \times U(1)_Y$), one always have 
an extra singlet scalar $s$ which couples to the hidden sector DM particles at 
renormalizable  level. And the mixing of  this singlet scalar $s$ with the SM Higgs $h$
thermalize the hidden sector DMs.  Due to the mixing between the $h$ and $s$,  there are
always at least two neutral Higgslike scalar bosons, which are mixtures of the SM Higgs 
$h$ (remnant of the $SU(2)_L$ doublet Higgs field) and a singlet scalar  $s$.  
The singlet scalar $s$ not only thermalizes the hidden sector DM efficiently, but also 
improves the stability of the EW vacuum up to Planck scale \cite{Baek:2012uj}, 
unlike the SM case.  It is important to realize that one can not integrate out the singlet 
scalar $s$  simply assuming that $s$ is very heavy.   The mixing should be
taken into account properly, the discussion of which can be found in Sec.~3.3 of 
Ref.~\cite{Baek:2012se}.  Furthermore, the usual effective Lagrangian based on 
$SU(3)_C \times U(1)_Y$  is not proper approach when one attempts to deduce 
information on new light particles close to the observed 125 GeV resonance  
and related underlying physics. 

In fact, an extra neutral scalar boson does appear in many interesting extensions of the SM,  
and it generally mixes with the SM Higgs boson unless the mixing is forbidden by some 
exact symmetry.  Introducing an extra singlet scalar boson is not only the simplest extension 
of the SM in terms of the number of new degrees of freedom, but also has various virtues 
of leaving  the $\delta \rho$ parameter intact at the tree level, and improving 
the vacuum stability \cite{EliasMiro:2012ay} as well as helping achieve the stronger first order 
electroweak phase transition \cite{Profumo:2007}. Also, SM singlet scalars can be  charged under 
a new gauge symmetry spntaneously breaking the symmety or playing an role of cold dark matter (CDM).
Furthermore, during the last few years or so, there have been a lot of works which considered singlet scalars coupling to extra 
vectorlike fermions or new charged vector bosons in order to enhance the diphoton rate of 125GeV resonance.   Most models involve the extra singlet scalar that mixes with the SM 
Higgs boson as described in Sec.~\ref{sec:models}.

Let us list some example BSM's (with a few referecens) where at least one neutral scalar boson appears and 
mixes with the SM Higgs boson $h$: 
\begin{itemize}
\item Pure singlet extension of the SM \cite{EliasMiro:2012ay,Schabinger:2005ei,Profumo:2007,O'Connell:2006wi,Barger:2007}
\item Hidden (dark) sector DM with hidden (dark) gauge symmetry \cite{Hur:2007uz,Ko:2008ug,Kim:2008pp,Hambye:2008bq,Gopalakrishna:2009yz,Feng:2009mn,Hur:2011sv,Baek:2011aa,Baek:2012uj,Baek:2012se,Baek:2013qwa}
\item Dilaton or radion \cite{Coleman:1974hr,Goldberger:1999uk,Giudice:2000av,Bae:2000pk,Cheung:2000rw,Goldberger:2007zk,Vecchi:2010gj,Appelquist:2010gy,Barger:2011hu,Cheung:2011nv,Matsuzaki:2012xx,Chacko:2012vm,ko_jung}
\item Enhancing diphoton rate from vector-like(VL) fermions or new charged vector bosons \cite{Carena:2012xa,Gunion:2012gc,voloshin,Batell:2012mj,Gunion:2012he,Moreau:2012da,Batell:2012zw,Abe:2012fb,Schwaller:2013hqa}
\end{itemize}

Determination of the Higgs couplings in the presence of a (light) singlet scalar boson 
has not been discussed properly in the recent literature (but, for some earlier attempts, see Refs.\cite{Carmi:2012yp,Bertolini:2012gu,Cheung:2013bn}).  In fact,  in most phenomenological analyses of 125 GeV boson, 
the usual approach actually does not distinguish the 125 GeV boson  discovered at the 
LHC being  the pure SM Higgs boson or a mixture with a singlet scalar boson for the 
reasons described in Sec.~\ref{sec:eff-lag}.

It is the purpose of this paper  to introduce a proper methodology for analyzing the LHC 
Higgs data in the presence of new physics that affects the SM Higgs couplings either 
directly or indirectly through the mixing of the SM Higgs boson with a singlet scalar that 
can couple to new physics. 

\begin{figure}
\begin{center}
\includegraphics[scale=0.35]{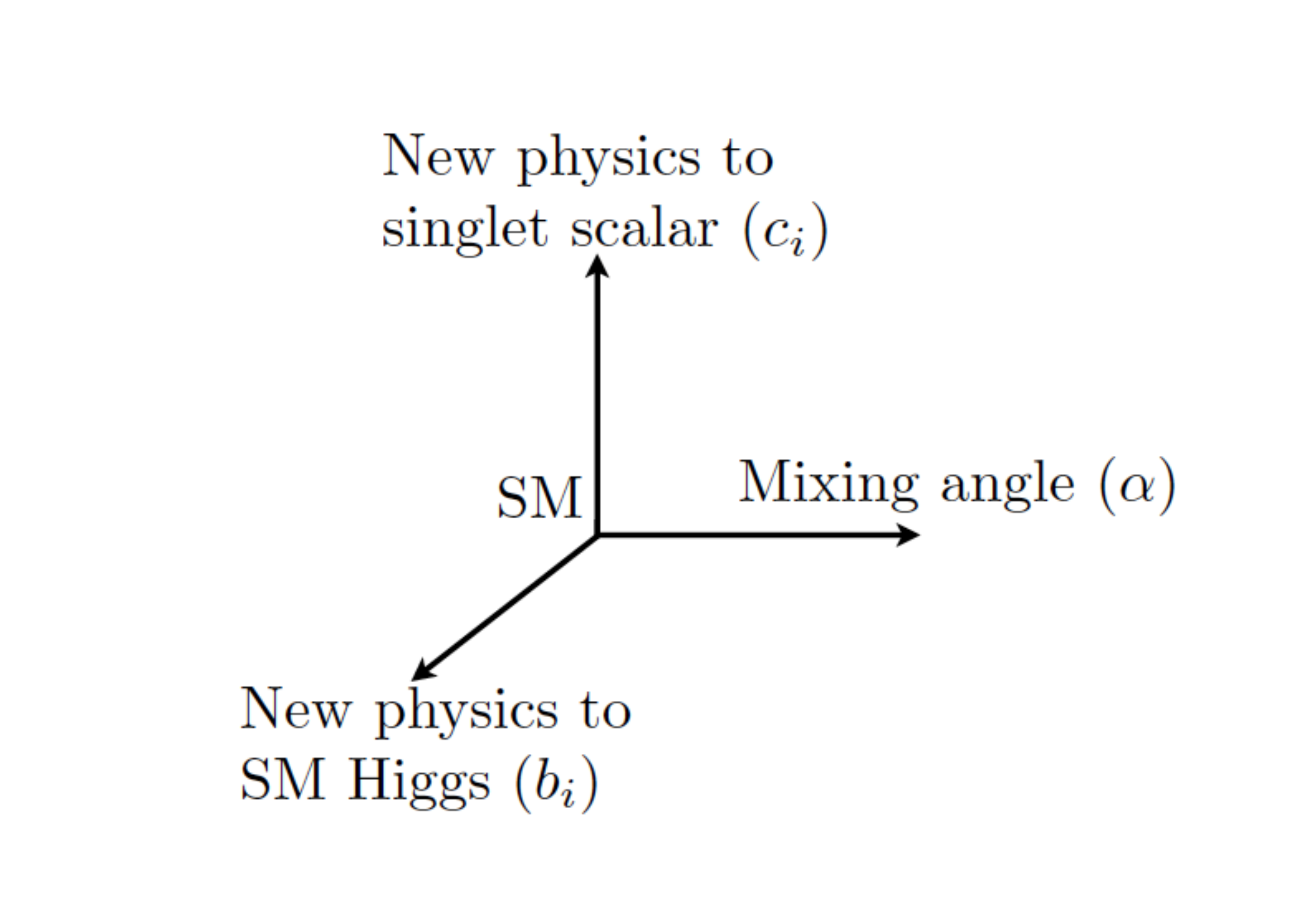}
\caption{\label{fignonfrag}
The field space diagram representing three independent ways that we can modify 
the properties of the SM Higgs boson. Parameters $b_i$, $c_i$ and $\alpha$ are
 discussed in text.}
\end{center} \label{fig:3axis}
\end{figure}

To this end, instead of parameterizing BSM effects on the mass-eigenstate of 125GeV resonance directly,  we introduce separate parameterizations for each \emph{interaction 
eigenstates of SM Higgs ($h$) and a  singlet scalar boson ($s$), and their mixing through 
the mixing angle $\alpha$. }    In other words, we introduce
\beq
\alpha, \quad b_i, \quad c_i
\eeq
for the mixing angle, corrections to SM Higgs couplings and the singlet 
scalar couplings to the SM fields $i = WW,ZZ, gg,\gamma\gamma, Z\gamma, f\bar{f}$, 
respectively (see Sec.2.1 and 2.2 for detail).   
With these parameterizations, physical interpretation of data and global fits will be different 
from usual  approaches where only final modified Higgs couplings can be measured.  
Usefulness of our approach is clear; one can directly extract information of Lagrangian
parameters and thus possibly underlying physics responsible for them. This parameterization also 
makes it clear that signal strengths can be modified from several independent sources  
simultaneously so that it is difficult to understand the origin  of modification without considering a proper Lagrangian; see \Fig{fig:3axis} for illustration of this important statement.   
By the same reason, data may be parameterized by sometimes redundant parameters 
implying not only that no unique solution of global fit may be found, but also that
some parameters maybe  hard to be constrained. 
We discuss how one can handle this difficulty, and how one can still deduce upper bound 
on non-standard branching ratio, which often causes this difficulty, in a generic way.

This framework is versatile enough to capture various interesting models of singlet scalar 
bosons ubiquitous in BSMs (as well as the new physics scenarios without a singlet scalar 
boson). Each case we consider in this paper has corresponding models and motivations. 
We also discuss how the fit results on these parameters can further be used to constrain 
new particles that may exist nearby  electroweak scale.  
This will prove the usefulness and the prospects of our approach.

This paper is organized as follows. In Sec.~\ref{sec:eff-lag}, we introduce effective Lagrangians  
useful to analyze the Higgs properties in the presence of a possible mixing between 
the SM Higgs boson and a singlet scalar boson nearby, including the cases where the 
singlet scalar boson couples to new particles such as vectorlike fermions or new 
charged vector bosons or scalar bosons.  In Sec.~\ref{sec:models}, we list a number of BSM's 
which have extra SM singlet scalar fields.  Then in Sec.~\ref{sec:fitting}, we explain our fit procedure 
and the LHC data we use. In Sec.~\ref{sec:sing-fit}, we perform global fits to the LHC Higgs data under 
various assumptions, encompassing all the models described in Sec.~\ref{sec:models}.  
In Sec.~\ref{sec:sing-near}, the fit results are combined with other collider searches of  Higgs-like resonances 
to constrain another scalar boson that may be present nearby electroweak scale.  
Finally the results are summarized in Sec.~\ref{sec:conc}. 

\section{Effective Lagrangian for the SM Higgs and a singlet scalar bosons} \label{sec:eff-lag}

In order to extract the Higgs couplings to the SM fermions and gauge bosons in a model independent fashion, it is useful to construct the most general effective Lagrangian 
describing the interactions between the SM Higgs boson $h$ or a singlet scalar boson $s$
with the SM fields such as $b, t$, $W^\pm , Z^0$, $\gamma$ and $g$.   
In this section, we show the effective Lagrangian for this purpose up to dim-6 operators.
We impose the full SM local gauge symmetry $SU(3)_c \times SU(2)_L \times U(1)_Y$ 
to the effective Lagrangian, not its unbroken subgroup $SU(3)_c \times U(1)_{\rm em}$. 
By doing so, we can separate two different sources of the modified  Higgs properties:  
one from direct couplings of new particles to the SM Higgs boson ($b_i \neq 1$ in Fig.~\ref{fig:3axis}), 
and the other from the mixing with a singlet scalar boson ($\alpha \neq 0$ in Fig.~\ref{fig:3axis}). 
There could be new particles that have gauge invariant renormalizable couplings to the 
singlet scalar $s$ ($c_i \neq 0$ in Fig.~\ref{fig:3axis}),  but not to the SM Higgs boson $h$.  
Therefore  studying the Higgs properties  in the 3-dimensional space (ignoring the   dimensionality associated with the index $i$)  as depicted  in Fig.~\ref{fig:3axis} can be justified,  
and its importance could be appreciated. 

\subsection{Effective Lagrangian for the SM Higgs boson $h$}

Let us assume that the SM Higgs boson couplings are modified due to some new physics  
effects even without the mixing with a singlet scalar $s(x)$.  This could happen if there are additional 
sequential or mirror  fermions (chiral), or extra inert scalar doublet, for example.
Integrating out the new heavy particles, one can construct the effective Lagrangian up to dim-5 and dim-6 
operators, all of which have been identified by Buchm\"{u}ller and Wyler sometime ago~\cite{Buchmuller:1985jz}.  We do not reproduce all the operators involving  the Higgs fields, but list only some of them just for  illustration:
\[H^\dagger H ~G_{\mu\nu}^a G^{a\mu\nu} , \ \ 
( H^\dagger D_\mu H ) ( H D^\mu H^\dagger ) , \ \
H^\dagger H ~\overline{Q_{3L}} \widetilde{H} t_R , \ \ 
\]
relegating the complete list to the original paper~\cite{Buchmuller:1985jz}. 

Expanding the Higgs field in the effective Lagrangian constructed by Buchmuller and Wyler 
around the EW vacuum with 
\[
H (x) = \left( \begin{array}{c} 
                    0 \\  v + h(x) \end{array} 
                    \right)  \ ,
\]
we obtain the following effective operators of \emph{interaction eigenstate} $h(x)$ field 
upto dim-6:
\begin{eqnarray}
- {\cal L}_{\rm h, int} & = &  \sum_f m_f \left\{ b_f \frac{h}{v} + \frac{1}{2} b_f^{'} \left( \frac{h}{v} \right)^2 
 \right\} \bar{f} f 
\nonumber  \\
& - & \left\{ 2 b_W \frac{h}{v} + b_W^{'} \left(\frac{h}{v}\right)^2 \right\}  m_W^2 W_\mu^+ W^{- \mu} 
-\left\{ b_Z \frac{h}{v} + \frac{1}{2} b_Z^{'} \left( \frac{h}{v} \right)^2 \right\} m_Z^2 Z_\mu Z^\mu 
\nonumber  \\
& + & 
\frac{\alpha}{8 \pi } r_{\rm sm}^\gamma \left\{ b_\gamma \frac{h}{v} + \frac{1}{2} b_\gamma^{'} 
\left( \frac{h}{v} \right)^2 \right\}   F_{\mu\nu} F^{\mu\nu}  
+ \frac{\alpha_s}{16 \pi} r_{\rm sm}^g \left\{ b_g  \frac{h}{v} + \frac{1}{2} b_g^{'} \left( \frac{h}{v} 
\right)^2 \right\}  G_{\mu\nu}^a G^{a\mu\nu}  
\nonumber 
\\
& + &  \frac{\alpha_2}{\pi} \left\{ 2 b_{dW} \frac{h}{v} + b_{dW'} \left( \frac{h}{v} \right)^2 \right\}  
W^+_{\mu\nu} W^{- \mu\nu} 
+ \frac{\alpha_2}{\pi} \left\{ 2b_{dZ} \frac{h}{v} + b_{dZ'} \left( \frac{h}{v} \right)^2 \right\} 
Z_{\mu\nu} Z^{\mu\nu} 
\nonumber
\\
& + &  \frac{\alpha_2}{\pi} \left\{ 2 \widetilde{b_{dW}} \frac{h}{v} + \widetilde{b_{dW'}} \left( \frac{h}{v} \right)^2 \right\} W^+_{\mu\nu} \widetilde{W^{- \mu\nu}}
+ \frac{\alpha_2}{\pi} \left\{ 2 \widetilde{b_{dZ}} \frac{h}{v} + \widetilde{b_{dZ'}} \left( \frac{h}{v} \right)^2 \right\} Z_{\mu\nu} \widetilde{Z^{\mu\nu}} 
\nonumber
\\
&+&  \frac{\alpha}{\pi} \left\{ 2 b_{Z\gamma} \frac{h}{v} + b_{Z\gamma'} \left( \frac{h}{v} \right)^2 \right\} F_{\mu\nu} Z^{\mu\nu} 
\label{eq:lag_higgs_h}
\end{eqnarray}
where $f$ in the first term of the Lagrangian denotes the SM fermions. 
The Higgs field $h(x)$ is defined {\it after the EWSB}: $H(x) = v + h(x)$, and {\it before any 
possible mixing with a singlet scalar $s$} which will be introduced shortly.  

Most of dim-6 operators lead to the definite relation, $b_i = b_i^{'}$, since they involve $H^\dagger H$ which 
yields $( v + h )^2$.  But this is not the case for $b_f $ and $b_f^{'}$. 
For example, the following operators ($q_L \equiv ( t_L , b_L )$), which are invariant under the full
SM gauge group $SU(3)_C \times SU(2)_L \times U(1)_Y$,  
\[
\overline{q_L} D_\mu b_R D^\mu H , \ \ \overline{q_L} D_\mu t_R D^\mu \widetilde{H} , \ \
\]
contribute to the $b_f \sim m_h^2 / \Lambda^2$, but not to $b_f^{'}$.    Thus the relation 
$b_f = b_f^{'}$ is no longer true for the Higgs couplings to the SM chiral fermions.

Modification to the SM Higgs Lagrangian is parameterized by multiplicative constants 
$b_i$ and $b_i^{'}$, and the SM is  recovered when all $b_i= b_i^{'} = 1$. We are interested in 
\beq
b_f,\, b_W,\, b_Z,\, b_\gamma,\, b_g
\eeq
among coefficients $b_i$ because these are most constrained by the current LHC data. 

Loop-induced couplings of the SM higgs to photons and gluons involve loop functions $r_{\rm sm}$ 
defined in the SM as
\begin{eqnarray}
r_{\rm sm}^\gamma & = &  A_1 (\tau_W) \+ N_c Q_t^2 A_{1/2} (\tau_t) 
\\
r_{\rm sm}^g & = &  A_{1/2}(\tau_t)
\end{eqnarray}
where we follow definitions of $A_1$ and $A_{1/2}$ as in Ref.\cite{Djouadi:2005gi}, and 
$\tau_i = m_h^2/4m_i^2$ and $v=246$ GeV. Loop effects of new physics is conveniently incorporated 
as additive shifts $\Delta b_\gamma$ and $\Delta b_g$ defined as
\beq
b_g  \=  b_t {\cal C}_t \+ \Delta b_g 
\ceq
b_\gamma \= b_W {\cal B}_W \+ b_t {\cal B}_t \+ \Delta b_\gamma 
\eeq
Note that $b_{g,\gamma}$ (and their $\Delta b$) are normalized to the corresponding SM couplings.  $b_t$ and $b_W$ parts describe effects from modification of top and $W$ 
boson couplings to Higgs that are  involved in loop diagrams. The relative loop-functions 
of $W$ boson and top quark for $m_h = 125$ GeV are given by
\beq
{\cal C}_t \= 1
\ceq
{\cal B}_{W} \= \frac{ A_1(\tau_W) }{ A_1(\tau_W) \+ N_c Q_t^2 A_{1/2}(\tau_t) } \, \simeq \, 1.283, 
\ceq
{\cal B}_{t} \= \frac{ N_c Q_t^2 A_{1/2}(\tau_t) }{ A_1(\tau_W) \+ N_c Q_t^2 A_{1/2}(\tau_t) } \, \simeq \, -0.283.
\label{eq:relloop}\eeq
These parameters $b_i$'s may have momentum(mass) dependence. We define these variables at 125GeV  relevant to the global fit to 125GeV resonance data. 

In most part of this paper, we work on the 125GeV  resonance, thus we can conveniently 
assume that $b_i$ do not have mass dependence.  However, for $b_g$ and $b_\gamma$ 
which are loop-induced couplings, we will discuss mass dependence in Sec.~\ref{sec:sing-near} when 
we study constraints on other particles. We also assume that these parameters are real. 
This assumption would be good as long as the loop diagram does not develop unitarity 
phase from the case  where the loop particles are on-shell.  
Considering various constraints on new charged or colored particles,  it would be 
reasonable to assume that there are no new charged or colored particles with mass 
less than $m_H /2 \simeq 63$GeV. 

In the presence of new particles with nonzero EW gauge charges (e.g., another Higgs 
doublet as in 2 Higgs doublet model, extra sequential fermions or mirror fermions), 
both tree level processes $h \rightarrow W^+ W^- , Z^0 Z^0$ and the loop process 
$h\rightarrow gg, \gamma\gamma$ can be modified, resulting in $b_V \neq 1$ and 
$b_\gamma \neq 1$ and $b_g \neq 1$.  Except for the 2HDM case, these new physics 
effects will appear at one loop level, and we would expect that 
\[
b_i \sim ``1" + \frac{g^2 m^2}{(4 \pi)^2 M^2} , \ \ {\rm or} \ \ ``1" + \frac{g^2 m^2}{M^2}
\]
where $m$ is the external SM particle mass, $M$ is the mass of new particles in 
the loop, and $g$ is the couplings between them. 

\subsection{Effective Lagrangian for a singlet scalar boson $s$}

As in the case of the effective Lagrangian of the SM Higgs field $H(x)$ up to dim-6, 
one can construct effective Lagrangian involving a singlet $S(x)$ and the SM fields up to dim-6, imposing the SM gauge symmetry $SU(3)_C \times SU(2)_L \times U(1)_Y$.   
Note that there are only a few operators describing interactions between $S$ and the 
SM Higgs boson at renormalizable level: 
\[
S ~H^\dagger H , \ \ S^2 ~H^\dagger H , \ \ 
\]
in addition to the singlet self couplings: $S^3$ and $S^4$, which lead to the modified self 
couplings of two Higgs-like scalar bosons $H_1$ and  $H_2$ after the EWSB and 
the mass mixing between $h$ and $s$, as described in  Sec. 2.3 below.

Interactions between the singlet scalar $S$ and the SM chiral fermions and the SM gauge 
bosons occur only at the nonrenormalizable level due to the full SM gauge symmetry\footnote{As discussed in Sec.~\ref{sec:intro}, the singlet scalar could have renormalizable interactions 
$S \bar{f} f$ with the SM fermions if we imposed only the unbroken part of the SM gauge 
symmetry. However this can lead to erroneous results as demonstrated in Refs.~\cite{Baek:2011aa,Baek:2012uj,Baek:2012se} in the context of Higgs portal DM models.}, 
$SU(3)_C \times SU(2)_L \times U(1)_Y$.   As an example, we list a few of them:   
\[
S~ G_{\mu\nu}^a G^{a\mu\nu} , \ \ S^2 ~G_{\mu\nu}^a G^{a\mu\nu} , \ \ 
S~ D_\mu H^\dagger D^\mu H ,  \ \ S^2 ~D_\mu H^\dagger D^\mu H , 
\]
\[
S ~\overline{Q_{3L}} \widetilde{H} t_R , \ \ , S^2 ~\overline{Q_{3L}} \widetilde{H} t_R ,  \ \ 
\]
etc.. We considered the most general Lagrangian without any symmetry  such as $Z_2$ 
symmetry under $S \rightarrow -S$ which is often invoked in the real singlet scalar DM 
models.  It would be a separate question what kind of new underlying physics would 
generate such dim-5 or dim-6 operators, which we don't address in this paper. 

The singlet  scalar field $S(x)$ may develop a nonzero VEV independent of the EWSB:
\[
S(x) = v_S + s(x). 
\]
Expanding around $v_S$, we define the physical singlet scalar $s(x)$ in the interaction basis. 
Then, the effective Lagrangian for the singlet \emph{interaction eigenstate} scalar boson $s$ 
could be written as
\begin{eqnarray}
- {\cal L}_{\rm s, int} & = &  \sum_f c_f \frac{m_f}{v} s \bar{f} f 
- \left\{ 2 c_W \frac{s}{v} + c_W^{'} \left(\frac{s}{v}\right)^2 \right\}  m_W^2 W_\mu^+ W^{- \mu} 
-\left\{ c_Z \frac{s}{v} + \frac{1}{2} c_Z^{'} \left( \frac{s}{v} \right)^2 \right\} m_Z^2 Z_\mu Z^\mu 
\nonumber  \\
& + & 
\frac{\alpha}{8 \pi } r_{\rm sm}^\gamma \left\{ c_\gamma \frac{s}{v} + \frac{1}{2} c_\gamma^{'} 
\left( \frac{s}{v} \right)^2 \right\}   F_{\mu\nu} F^{\mu\nu}  
+ \frac{\alpha_s}{16 \pi} r_{\rm sm}^g \left\{ c_g  \frac{s}{v} + \frac{1}{2} c_g^{'} \left( \frac{s}{v} 
\right)^2 \right\} G_{\mu\nu}^a G^{a\mu\nu}  
\\
& + &  \frac{\alpha_2}{\pi} \left\{ 2 c_{dW} \frac{s}{v} + c_{dW'} \left( \frac{s}{v} \right)^2 \right\} 
W^+_{\mu\nu} W^{- \mu\nu} 
+ \frac{\alpha_2}{\pi} \left\{ 2c_{dZ} \frac{s}{v} + c_{dZ'} \left( \frac{s}{v} \right)^2 \right\} 
Z_{\mu\nu} Z^{\mu\nu} 
\nonumber
\\
& + &  \frac{\alpha_2}{\pi} \left\{ 2 \widetilde{c_{dW}} \frac{s}{v} + \widetilde{c_{dW'}} \left( \frac{s}{v} \right)^2 \right\} W^+_{\mu\nu} \widetilde{W^{- \mu\nu}}
+ \frac{\alpha_2}{\pi} \left\{ 2 \widetilde{c_{dZ}} \frac{s}{v} + \widetilde{c_{dZ'}} \left( \frac{s}{v} \right)^2 \right\} Z_{\mu\nu} \widetilde{Z^{\mu\nu}} 
\nonumber
\\
&+&  \frac{\alpha}{\pi} \left\{ 2 c_{Z\gamma} \frac{s}{v} + c_{Z\gamma'} \left( \frac{s}{v} \right)^2 \right\} F_{\mu\nu} Z^{\mu\nu}  \quad  -\,{\cal L}_{nonSM}
\label{eq:lag_higgs_s}
\end{eqnarray}
The newly introduced couplings $c_i$'s  parameterize the couplings of $s$ to the SM 
particles in a similar way to the SM Higgs ($h$) couplings to the SM particles. 
The singlet interaction eigenstate $s(x)$ is defined {\it after the symmetry breaking 
due to possible nonzero VEV of a singlet scalar field $S(x)$ but before mixing with the SM Higgs field   $h$}. 
The last term ${\cal L}_{nonSM}$ represents possible interactions of the singlet scalar 
$s$ with non-SM particles such as dark matter in some dark matter models such as 
hidden sector dark matter models.  We do not specify this Lagrangian, but we will 
parameterize this effect by non-standard branching ratio in later sections.

Since all the couplings $c$'s are from nonrenormalizable interactions between the singlet scalar $S$
and the SM fields (except for the Higgs fields), one can assume that $c$'s are all suppressed
by heavy mass scale and/or the loop suppression factors:
\[
c_i \sim  ``0"  + \frac{g^2 m^2}{(4 \pi)^2 M^2} ,   \ \ ``0" + \frac{g^2 m^2}{M^2} ,
\]
where $m^2$ indicates the mass scales of the external SM particles or the singlet scalar.  
Therefore the natural scale for the $c_i$'s would be parametrically suppressed relative to ``1''.  

Also, these parameters $c_i$'s may have momentum (or mass) dependence. 
We define these parameters at 125GeV which is relevant to global fit to 125GeV 
resonance data. In most part of this paper, we work on 125GeV resonance, thus we can 
conveniently assume that $c_i$ do not have mass dependence. However, for $c_g$ 
and $c_\gamma$ which are loop-induced couplings, we will discuss mass dependence in 
Sec.~\ref{sec:sing-near} when we study constraints on other particles. We also assume that these 
parameters are real as before in case of $b_g$ and $b_\gamma$.

Note that $c_{f} = c_{V} = 0$ (with $V=WW,ZZ$) at renormalizable level in models with 
a singlet scalar boson because of the local gauge symmetry of the SM 
$SU(3)_c \times SU(2)_L \times U(1)_Y$ forbids renormalizable couplings of the singlet 
scalar $s$ to the SM fields except for the $H^\dagger H$ operator.   
In case of the hidden sector dark matter models (or Higgs-portal DM models), 
the singlet $s$ can couple to the SM singlet particles such as hidden sector dark 
matters (Sec.~3.2), but not to the SM fermions or weak gauge bosons at renormalizable 
level.  However  $c_{f}$ and $c_{V} $ could be nonzero if $s$ is a dilaton in spontaneously broken scale symmetric scenario or the radion in the Randall-Sundrum scenario 
where all the SM fields are on TeV brane (see Sec.~3.3). Unlike the couplings $c_f$ 
or $c_V$,  the singlet couplings to the $\gamma\gamma ,\,  gg,\, Z \gamma$ can be  nonzero 
if there are mixings between the SM fermions and extra singlet fermions (see Sec.~3.4), 
of if there are an extra charged vector bosons in the theory which get massive through 
new Higgs fields (Sec.~3.5).   In Sec.~3.6, we give an explicit example with (colored and) 
charged scalar fields,  in which contributes  $b_g$  and $b_\gamma$ can be modified from 
the SM values $1$,  and also $c_g$ and $c_\gamma$ can be generated.

\subsection{Mixing between $h$ and $s$ and physical amplitudes}

In general, there would be a mass mixing between $h(x)$ and $s(x)$ after $H(x)$ and 
$S(x)$ develop nonzero VEV's. Relevant nonlinear interactions among them are given by
\begin{eqnarray}
- {\cal L}_{\rm bilinear} & = & \frac{1}{2} m_h^2 h^2 + \frac{1}{2} m_s^2 s^2 
+ m_{hs}^2 h s  \ ,
\\
- {\cal L}_{\rm scalar int} & = &  \frac{a_{3,0}}{3 !} h^3 + \frac{a_{4,0}}{4!} h^4 
+ \frac{a_{2,1}}{2 !} h^2 s + \frac{a_{3,1}}{3 !} h^3 s 
\nonumber  \\
& + & \frac{a_{1,2}}{2 !} h s^2 + \frac{a_{2,2}}{( 2! )^2} s^2 h^2 + 
\frac{a_{0,3}}{3 !} s^3 + \frac{a_{1,3}}{3 !} s^3 h  \ .
\end{eqnarray}
We included the scalar self couplings for completeness, although they are not relevant
to the discussions in this work. Let us  parameterize the mixing effects by a mixing angle $\alpha$ 
which defines the physical mass eigenstates $H_1$ and $H_2$ as 
\begin{eqnarray}
H_1 & = & h \cos \alpha - s \sin \alpha 
\\
H_2 & = & h \sin \alpha + s \cos \alpha
\end{eqnarray}
where we conveniently denote 125GeV resonance by $H_1$ although it can be heavier 
or lighter than $H_2$. Their partial widths to the SM particles $F (\neq H_{i=1,2})$\footnote{Note that in our definition $F$ denotes  the SM fields only, so that interaction eigenstate $s$ 
does not have couplings to $F$ except for the case $F = h$.} are written as 
\begin{eqnarray}
\left. \frac{\Gamma(H_1 \to F)}{\Gamma(h \to F)_{SM}} \right|_{m_{H_1}} & = &  \left( b_{F} \cos\alpha 
- c_{F} \sin\alpha \right)^2  
\\
\left. \frac{\Gamma(H_2 \to F)}{\Gamma(h \to F)_{SM}} \right|_{m_{H_2}} & = &  \left( c_{F} \cos\alpha 
+ b_{F} \sin\alpha \right)^2 
\end{eqnarray}
Note that we normalize the decay widths of two physical scalar boson with respect to 
corresponding SM width at the mass of the scalar boson. We treat $b_i$ and $c_i$ are 
mass-independent if they are generated at tree-level; thus, their values fitted at 125GeV are  also applied to other mass 
region. We discuss how loop-induced couplings which are mass-dependent can be 
treated in Sec.~\ref{sec:sing-near} when we study constraints on other particles.

Another possible effect of mixing is that the heavier eigenstate can decay to the lighter 
one if it is kinematically allowed. We will parameterize this effect by introducing 
non-standard branching ratio in Sec.~\ref{sec:fitting}.

\subsection{Comparison with other approaches}

Before proceeding further, let us compare our approach with others. 
Most papers use the effective Lagrangian \Eq{eq:lag_higgs_h} as the starting point, assuming that 
the $h$ in \Eq{eq:lag_higgs_h} is the SM Higgs boson derived from the SM Higgs double $H$ 
and imposing the unbroken part of the SM gauge group, namely imposing only 
$SU(3)_C \times U(1)_{\rm em}$.  
There is nothing wrong about this, since it would be the most general effective 
Lagrangian up to dim-6 when we impose local $SU(3)_C \times U(1)_{\rm em}$ symmetry. 
However one has to be careful since the Higgs field $h$ in \Eq{eq:lag_higgs_h} with local 
$SU(3)_C \times U(1)_{\rm em}$ symmetry could be different from the genuine SM Higgs 
field, the remnant of the $SU(2)_L$ doublet scalar fields after EWSB.   If we consider 
local $SU(3)_C \times U(1)_{\rm em}$ symmetry, then the Higgs field in \Eq{eq:lag_higgs_h} could be a 
mixture of the SM Higgs field and any number of electrically neutral scalar fields, some of 
them could be EW singlets and others could carry nontrivial EW gauge charges.  
Therefore there is no way one can tell whether the  observed 125GeV boson is the SM 
Higgs boson or a mixture with a singlet  scalar boson  within the usual approach. 

In contrast, we are proposing to separate $h$ and $s$ in the effective Lagrangian 
from the beginning by their EW gauge quantum numbers. Therefore one can interpret the 
global fit results under various assumptions on the underlying new physics models and tell 
which models are favored and which are not.   At the moment, the data currently available 
is not good enough to constrain or exclude some BSM's definitely.  However in the future 
when more data is available with better information on the production channels,  
our approach would be useful for constraining various BSM's as well as verifying the SM 
Higgs scenario. 

\section{Models with an extra singlet scalar} \label{sec:models}

In this section, we consider a number of models where extra singlet scalar bosons 
appear in a natural manner.  The scalar $S(x)$ is assumed to be a singlet under 
the SM gauge group, but could be either neutral or charged under a new gauge group.
The hidden sector Higgs fields~\cite{Baek:2012se} or a new singlet scalar charged 
under $U(1)_H$ in the Type-II two Higgs doublet models with gauged Higgs flavor 
proposed in Ref.~\cite{Ko:2012hd} make good examples.  

In many cases, the singlet scalar $S$ develops nonzero VEV $v_S$, 
the physical scalar $s(x)$ is defined as  a fluctuation around a nonzero VEV of 
the original singlet field $S$: 
\[
S(x) = v_S + s(x). 
\]
In this case, $s$ can not have tree level couplings to the SM fermions or weak gauge 
bosons: $c_V = c_f = 0$, but it may have couplings to new charged particles such 
as vector-like(VL) fermions (either colored fermion $Q$'s or colorless leptons $L$'s) 
\cite{voloshin,Batell:2012zw} 
or new charged vector bosons such as $W^{' \pm}$ \cite{Carena:2012xa,Abe:2012fb}.  
In either case, these new particles ($Q, L, W^{' \pm}$'s)  can modify the Higgs 
phenomenology only through the mass mixing  between $h$ and $s$, since they couple 
only to the singlet $s$ but not to the SM Higgs boson $h$.  They will induce the nonzero 
amplitudes for $s \rightarrow \gamma \gamma$ and $s\rightarrow g g$ through loop
diagrams, so that $c_\gamma \neq 0$ for all of them and $c_g \neq 0$ for $Q$'s only.
We assume these new charged particles are  heavy enough that the 125 GeV 
resonance discovered at the LHC cannot decay into those particles.  Then it would be 
possible to use the effective Lagrangian for $H (125) \rightarrow \gamma \gamma$ 
with the couplings $c_\gamma$ and $c_g$ being real. 

\subsection{The singlet extension of the SM}

The simplest extension of the SM Higgs sector is to add a singlet scalar $S$ to the 
SM Lagrangian: 
\begin{equation}
 {\cal L} = {\cal L}_{\rm SM} +  {\cal L}_{\rm new},
\end{equation}
where 
\begin{equation}
{\cal L}_{\rm new} = {1 \over 2} (\partial_\mu S \partial^\mu S - m_S^2 S^2) 
-\mu_S^3 S - {\mu_S^\prime \over 3} S^3  - {\lambda_S \over 4} S^4 
- \mu_{HS} S H^\dag H -{\lambda_{HS} \over 2} S^2 H^\dag H,
\end{equation}
This model has been studied in various context in 
~\cite{O'Connell:2006wi,Profumo:2007,Barger:2007}.   After the singlet $S(x)$ develops 
a nonzero VEV $v_S$, a physical singlet scalar $s(x)$ is defined as 
\[
S(x) = v_S + s(x). 
\]
In this case, $b_F = c_F = 0$ with $F = g, W, Z, \gamma , b , \tau$, 
and the SM Higgs signal is diluted by the mixing angle $\alpha$.  
Only possible modification for the Higgs phenomenology is in the Higgs self coupling, 
which is beyond the scope of this paper however.

\subsection{Hidden sector DM models}

In this case, the singlet $s$ comes either from the messengers between the SM 
and the hidden sectors, or from the remnant of the hidden sector Higgs fields after 
the spontaneous breaking of hidden sector local gauge symmetry.  

\subsubsection{Singlet fermion DM}

Let us consider a singlet fermion DM model with a singlet Dirac dark matter $\psi$ and a 
singlet scalar $S$ as a messenger between the SM sector and the dark matter sector. 
Then, the model Lagrangian has 3 pieces, the hidden sector and Higgs portal terms 
in addition to the SM Lagrangian~\cite{Baek:2011aa,Baek:2012uj}: 
\begin{equation}
 {\cal L} = {\cal L}_{\rm SM} + {\cal L}_{\rm hidden} + {\cal L}_{\rm portal},
\end{equation}
where 
\begin{eqnarray}
{\cal L}_{\rm hidden} &=& {\cal L}_S + {\cal L}_\psi - \lambda S \overline{\psi} \psi, 
\nonumber  \\
{\cal L}_{\rm portal} &=& - \mu_{HS} S H^\dag H -{\lambda_{HS} \over 2} S^2 H^\dag H,
\end{eqnarray}
with 
\begin{eqnarray}
{\cal L}_S &=&
{1 \over 2} (\partial_\mu S \partial^\mu S - m_S^2 S^2) 
-\mu_S^3 S - {\mu_S^\prime \over 3} S^3  - {\lambda_S \over 4} S^4, \nonumber \\
{\cal L}_\psi &=&
\overline{\psi} ( i \not \partial - m_{\psi_0} ) \psi.
\end{eqnarray}
It is important to introduce a singlet scalar $S$ as a messenger in order to have correct thermal relic density of singlet fermion DM and to relax the stringent upper bounds on 
the Yukawa coupling $\lambda$ from direct detection experiments
~\cite{Baek:2011aa,Baek:2012uj}.  
The model without the dark sector, namely the SM plus an additional singlet scalar 
field has been studied in detail in ~\cite{Profumo:2007,Barger:2007}.
After the singlet $S(x)$ will develop VEV $v_S$, a physical singlet scalar $s(x)$ 
is defined as 
\[
S(x) = v_S + s(x). 
\]
The detailed study of this singlet fermion DM with Higgs portal has been presented 
in Ref.~\cite{Baek:2011aa,Baek:2012uj}. 

\subsubsection{Higgs portal Abelian vector DM}

For the case of hidden sector vector dark matter, one has to introduce 
a hidden sector Higgs field in order to generate the vector DM mass~\cite{Baek:2012se}.
Let us consider a vector boson dark matter, $X_\mu$, which is assumed to be
a gauge boson associated with Abelian dark gauge symmetry $U(1)_X$. 
The simplest model will be without any matter fields charged under $U(1)_X$ 
except for a complex scalar, $\Phi$ (a SM singlet), whose VEV will generate 
the mass for $X_\mu$ by the conventional Higgs mechanism:
\begin{eqnarray}
{\cal L}_{VDM} & = & - \frac{1}{4} X_{\mu\nu} X^{\mu\nu} +  
(D_\mu \Phi)^\dagger (D^\mu \Phi) 
- \frac{\lambda_\Phi}{4} \left( \Phi^\dagger  \Phi - \frac{v_\Phi^2}{2} \right)^2
\nonumber \\
& & - \lambda_{H\Phi} \left(H^\dagger H - \frac{v_H^2}{2}\right) 
\left(\Phi^\dagger \Phi - \frac{v_\Phi^2}{2}\right) \ ,
\end{eqnarray}
in addition to the SM lagrangian. The covariant derivative is defined  as 
\[
D_\mu \Phi = (\partial_\mu + i g_X Q_\Phi X_\mu) \Phi ,
\]
where $Q_\Phi \equiv Q_X(\Phi)$ is the $U(1)_X$ charge of $\Phi$.

Assuming that the $U(1)_X$-charged complex scalar $\Phi$ develops a nonzero VEV, 
$v_\Phi$, and thus breaks $U(1)_X$ spontaneously, 
\[
 \Phi  = \frac{1}{\sqrt{2}} \left( v_\Phi+ \varphi(x) \right) .
\]
Therefore the Abelian vector boson $X_\mu$ get mass $m_X^2 = g_X^2 Q_\Phi^2 v_\Phi^2$, 
and the hidden sector Higgs field (or dark Higgs field) $\varphi (x)$ will mix with 
the SM Higgs field $h(x)$ through Higgs portal of the $\lambda_{H\Phi}$ term.
The detailed phenomenogical analysis of this model is presented in Ref.~\cite{Baek:2012se}. 
Here, it would suffice to mention that this higgs portal vector DM model is a 
viable model for CDM and the EW vacuum is stable upto Planck scale without 
any more new fields.   
In this paper, we presented this model lagrangian as an example where a singlet 
scalar appears naturally, by identifying $\Phi (x) \rightarrow S(x)$ and 
$\varphi (x) \rightarrow s(x)$.

\subsubsection{Scalar DM with local $Z_2$ symmetry}

Let us assume the dark sector has a local $U(1)_X$ gauge which is spontaneously
broken into local $Z_2$ symmetry.  This can be achieved with two complex scalar 
fields $\phi_X$ and $X \equiv X_R + i X_I$ in the dark sector with the $U(1)_X$ 
charges equal to $2$ and $1$, respectively, in the following lagrangian~\cite{ko_park}: 
\begin{eqnarray}
{\cal L} & = & {\cal L}_{\rm SM} 
- \frac{1}{4} X_{\mu\nu} X^{\mu\nu} - \frac{1}{2} \epsilon X_{\mu\nu} B^{\mu\nu} 
+ D_\mu \phi_X D^\mu \phi_X - \frac{\lambda_X}{4} 
\left( \phi_X^\dagger \phi_X - v_{\phi}^2 \right)^2  
\nonumber \\
& + & D_\mu X^\dagger D^\mu X 
- m_X^2 X^\dagger X 
-  \frac{\lambda_X}{4} \left( X^\dagger X \right)^2  -  \left( \mu X^2 \phi^\dagger 
+  H.c. \right) 
\\
& - & \frac{\lambda_{XH}}{4} X^\dagger X H^\dagger H - 
\frac{\lambda_{\phi_X H}}{4} \phi_X^\dagger \phi_X H^\dagger H
- \frac{\lambda_{XH}}{4} X^\dagger X \phi_X^\dagger \phi_X
\nonumber 
\label{eq:model}
\end{eqnarray}
After the $U(1)_X$ symmetry breaking by nonzero $\langle \phi_X \rangle = v_\phi 
\neq 0$, the $\mu-$term generates 
\[
( X^2 + H.c.)  = 2 ( X_R^2 - X_I^2) 
\]
which lifts the mass degeneracy between $X_R$ and $X_I$.  
One could also consider fermion DM with local $Z_2$ in a similar manner.  
The detailed phenomenology of this model will be presented in Ref.~\cite{ko_park}.

For the purpose of this paper, it is enough to show that the model above is a well defined 
scalar DM model with local $Z_2$ symmetry stabilizing the CDM, and improves the usual  
scalar DM based on $Z_2$ symmetry $S \rightarrow -S$ \cite{Burgess:2000yq}:
\begin{equation}
{\cal L}_{\rm SDM} = \frac{1}{2} \partial_\mu S \partial^\mu S - \frac{1}{2} m_S^2 S^2 
- \frac{\lambda_{SH}}{2} S^2 H^\dagger H - \frac{\lambda_S}{4 !} S^4 \ .
\end{equation}
Note that the scalar CDM model with DM being stabilized by local $Z_2$ is much more 
complicated than the usual $Z_2$ scalar CDM models.  In particular, there appears a 
singlet scalar $s(x)$ that mixes with the SM Higgs boson after the hidden $U(1)_X$ 
symmetry breaking.   Therefore the Higgs phenomenology will be necessarily modified 
due to the mixing with the singlet scalar $s(x)$.

\subsubsection{Parametrizations of hidden sector DM models}

In all the cases described in the previous subsubsections, we end up with a 
singlet scalar $s$ that mixes with the SM Higgs boson.  Therefore $c_F = 0$ for $F$ 
being the SM fields $F=V,f,g,\gamma$,  and only $c_\chi$ and $c_{h^2}$  are nonzero,  
where $\chi$ denotes CDM field (scalar, fermion or vector CDM in the hidden sector with 
Higgs portal).  In this case, there would be  universal signal reductions  in all the decay 
channels of $H_i$, whether the $H_i \rightarrow \chi \chi$ is kinematically open or not. 
Also note that nonzero $c_\chi$ can lead to invisible decays of $H_i$ after the mixing, 
if kinematically allowed.  These two simple predictions of the hidden sector DM models
can be tested at the LHC if more data on the Higgs boson is accumulated  in the future.

\subsection{Dilaton and Radion}

Both dilaton in technicolor models with approximate scale symmetry and the radion 
in the RS scenario couple to the scale anomaly of the theory, which is nothing but 
the trace of energy-momentum tensor $T_\mu^\mu$.   Usually it is assumed that 
\begin{equation}
T_\mu^\mu = 2\mu_H^2 H^\dagger H + \sum_f  m_f \bar{f} f - 2m_W^2 W^+ W^-  - 
m_Z^2 Z_\mu Z^\mu + \frac{\beta}{g}
 G_{\mu\nu} G^{\mu\nu}  
\label{eq:tmn-old}
\end{equation}
namely the trace of the energy-momentum tensor after EWSB and only the unbroken 
subgroup of the SM gauge symmetry (namely $SU(3)_C \times U(1)_{\rm em}$) 
is imposed.  This ansatz might be  fine if one were interested in the case without 
fundamental Higgs such as technicolor models or other models with dynamical EWSB. 

On the other hand, if the dilaton or the radion appears before the EWSB, it would be
more appropriate to impose  the full SM gauge symmetry for the dilaton/radion couplings 
to the SM fields. If we impose the full $SU(3)_C \times SU(2)_L \times U(1)_Y$ gauge 
symmetry, the SM Lagrangian has only one operator that breaks the scale symmetry 
explicitly, namely the Higgs boson mass term:
\begin{equation}
T_\mu^\mu = 2\mu_H^2 H^\dagger H + \frac{\beta}{g}
 G_{\mu\nu} G^{\mu\nu}  
\label{eq:tmn-new}
\end{equation}
Therefore, the dilaton will couple to the $H^\dagger H$ operator only at classical level, 
and to the scale anomaly at quantum level.   It turns out that two prescriptions for 
$T_\mu^\mu$, \Eq{eq:tmn-old} vs. \Eq{eq:tmn-new}, have vastly different phenomenological consequence 
in the Higgs-dilaton sector~\cite{ko_jung}.  Also note that dilaton couplings have extra 
pieces which are different from other singlet scalar cases,  namely nonzero $b_{dW}$  
and $b_{dZ}$.  Phenomenological analysis with the SM Higgs mixing with dilation or 
radion will be presented in detail elsewhere \cite{ko_jung}, and not covered in this paper. 

\subsection{Extra vectorlike fermions for enhanced $H\rightarrow \gamma\gamma$}

In this case, there could be an extra singlet scalar field in order to have 
renormalizable interactions between the  vector like fermions. 

\subsubsection{$SU(2)_L$ singlet vectorlike lepton}

If the vectorlike fermions are colorless $SU(2)_L$ singlets ($S_L^-$ and $S_R^-$)  
with electric charge $Q_e = - 1 = Y$,  it can not directly couple to the SM Higgs doublet. 
Therefore one has to introduce a singlet scalar field $S$: 
\begin{equation}
{\cal L} = \overline{S_L^-} i \not D S_L^- + \overline{S_R^-} i \not D  S_R^- 
-  \left\{ \overline{S_L^-} \left( m_S + \lambda S \right) S_R^- + 
y_{Si}~ \overline{l_{Li}} H S_R^- + H.c. \right\}
\end{equation}
Note that $S^-$ will mix with the $e_{Rj}$ after EWSB, and $S^-$ will decay 
to $h l^-$ through $y_{Si}$ couplings, and its contribution to $H\rightarrow \gamma\gamma$ 
is suppressed by the Yuaka coupling $y_{Si}$ that is presumably small.   
Also $S^-$ can not carry color charge, since we cannot introduce $y_{Si}$ terms 
which make $S^-$ decay. 

Note that the vectorlike charged scalar $S^\pm$ will generate 
$s\rightarrow \gamma\gamma$ and $s\rightarrow Z\gamma$ at one loop level. 
Therefore both $c_\gamma$ and $c_{Z\gamma}$ are nonzero, and they would affect 
$H\rightarrow \gamma\gamma$ and  $H \rightarrow Z\gamma$ through the mixing 
between the SM Higgs boson and the singlet scalar $s$.

\subsubsection{$SU(2)_L$ doublet plus singlet vectorlike leptons}

If we consider just one vectorlike leptons $L$ and $R$ in the $SU(2)_L$ doublet representation 
with $Y = -1/2$, 
\begin{equation}
L = ( N^0 , E^- )_L^T , \ \ \ R = ( N^0 , E^- )_R^T 
\end{equation}
one has to introduce two singlet fermions $S_L^-$ and $S_R^-$ in order to write 
couplings to SM Higgs field~\cite{voloshin,Batell:2012zw}:
\begin{equation}
{\cal L}_{\rm mass} = - m_1 \overline{S_L} S_R - m_2 \overline{L} R - \sqrt{2} y_{12} 
\overline{S_L} H^\dagger R - y_{21} \overline{L} H S_R + H.c. 
\end{equation}
The model contains one neutral Dirac lepton $N^0$ and two Dirac leptons $S^-$ and $E^-$ 
with electric charge $Q_e = -1$. The mass matrix of these new leptons is written as 
\begin{equation}
{\cal M} (v) = \left( \begin{array}{cc}
                              m_1 & y_{12} v \\
                              y_{21} v & m_2 
                              \end{array}
                              \right)
\end{equation}
in the $(S^- , E^- )$ basis.  The new Yukawa couplings $y_{12}$ and $y_{21}$ can be complex
in general, and there could be CP violation in $H\rightarrow \gamma\gamma$ as pointed out 
by M. Voloshin in Ref.~\cite{voloshin}. 

In this model, however, the neutral component $N^0$ which would be absolutely stable and 
could overclose the universe. This trouble can be resolved if we introduce a real singlet scalar 
$S$ as in the previous subsubsection, adding the following Yukawa couplings to the above 
lagrangian:  
\begin{equation}
\delta {\cal L}_{\rm mass} = - \lambda_{S} S \overline{L} R  
- \lambda_{S}^{'} S \overline{S_L} S_R + H.c.
\end{equation}
After $S(x)$ gets a nonzero VEV, $s$ will mix with the   
SM Higgs boson, and $N^0 \overline{N^0}$ can pair annihilate into the SM fields through 
Higgs portal: 
\[
N^0 \overline{N^0} \rightarrow s^* \rightarrow h^* 
\rightarrow {\rm SM \ \ particles}
\]
and can be thermalized. 

In this model for vectorlike leptons, both $c_\gamma$ and $c_{Z\gamma}$ can be 
nonzero, depending on the SM quantum numbers\footnote{For vectorlike quarks, both $c_g$ could be also nonzero in addition to 
$c_\gamma$ and  $c_{Z\gamma}$.}.
Also $c_{f'} \neq 0$, where $f^{'}$ denotes the vector like fermions.
The mixing due to the SM charged lepton and extra $S^-$ can induce nonzero 
$b_\gamma$  and $b_{Z\gamma}$, proportional to $y_{Si}$, but we ignore it here for 
simplicity assuming $y_{Si}$ is small enough.

\subsection{Extra charged vector bosons}

In this subsection, let us describe some models with extra charged vector bosons
that can contribute to $H\rightarrow \gamma \gamma$. 

In Ref.~\cite{Carena:2012xa,Abe:2012fb}, extra charged vector bosons $W^{'}$ in the $SU(2)_L$ triplet 
representation were considered as a possible  explanation for the enhancement 
of $H \rightarrow \gamma \gamma$ using the effective operator:
\begin{equation}
{\cal O_{W'}} = \frac{1}{2} c_{W'} g^2 H^\dagger H W_{~\mu}^{'+} W^{' - \mu} 
\end{equation} 
with $m_{W'}^2 = m_0^2 + c_{W'} g^2 v^2$.  However, this operator is not really 
renormalizable, since we need to introduce a new Higgs field $H^{'}$ that gives 
mass to the $W^{'}$. After symmetry breaking from 
$\langle H^{'} \rangle = v^{'} \neq 0$, there may be a remnant of symmetry breaking
in terms of a SM singlet scalar $s$ that couples to $W_{~\mu}^{'+} W^{' - \mu}$. 
In principle, these $W^{' \pm}$ can mix with the SM weak gauge boson $W^\pm$ 
in general. For example, this is well known in the $SU(2)_R$ extension of the SM.
In this paper we will ignore this mixing between the  $W^{' \pm}$ and $W^{\pm}$
for simplicity. 

Another simple case would be the $SU(2)_R$ extension of the SM, where there appear 
3 more gauge bosons, $W^\pm_R$ and $W^0_R$, that would mix with the SM 
weak gauge bosons $W^\pm$ and $Z^0$.  In this case, there will be a singlet scalar
$s$ that couples to $W_R^\pm$ after $SU(2)_R$ symmetry breaking. Also this $s$
will mix with the SM Higgs boson. 

We assume that $W'$ is significantly heavier than the SM $W$ boson
and the physical Higgs boson of mass 125 GeV.  In this case, all the $c_F$'s  
are zero, except for $c_\gamma$ due to the $s \rightarrow \gamma \gamma$
through $W'$ loop and $c_{h^2}$.

\subsection{Extra charged scalar bosons}

One can consider extra charged scalar bosons\footnote{Here, charged scalar boson 
means scalar bosons in nontrivial representation of the SM gauge group.}. 
In this case, it is not mandatory for an extra singlet scalar to appear in the models, 
but we can introduce them if one wishes, with the following operators:
\[
H^\dagger H \phi_a^\dagger \phi^a , \ \ S \phi_a^\dagger \phi^a , 
\  \ S^2 \phi_a^\dagger \phi^a ,
\] 
where $\phi_a$ are new scalar bosons with nonzero electric charge and/or color charge.
Then the SM Higgs properties  can modified only by the higher dimensional 
operators through modified $b_F$ with $F = g$ or $\gamma$ depending on the 
SM charges of the scalar bosons, where as $c_g$ and/or $c_\gamma$ can be nonzero.

\subsection{Summary}

We assumed that the extra singlet scalar $s$ is an EW singlet, and does not participate 
in EWSB. The resulting possible new physics contributions are parametrized in terms
of $c_i$'s, and we tabulate the nonzero $c$'s in Table~\ref{tab:cici}.
If the singlet $s$ was coming from another Higgs doublet which break EW symmetry, 
we should have introduced another parameters (analogous to $\tan\beta$ in 2HDMs) 
in addition to the mixing angle $\alpha$, and the analysis will be more involved than 
presented in this paper.  There are a number of analysis within 2 HDMs in the literature
~\cite{Aoki:2009ha,Branco:2011iw,Ferreira:2011aa,Barroso:2012wz,Baak:2012kk,Chang:2012ve,Ferreira:2012nv,Chiang:2013ixa,Barroso:2013zxa,Swiezewska:2012eh,Krawczyk:2013gia,Lee:2013fda,Altmannshofer:2012ar}, 
and we do not pursue this possibility in this paper. 

\begin{table}[th]
\begin{center}
\begin{tabular}{c|c}
\hline \hline
Model   &     Nonzero  $c_F$'s     
\\
\hline \hline
Pure Singlet Extension &   $c_{h^2} $ 
\\
Hidden Sector DM &   $c_\chi $,$c_{h^2} $ 
\\
Dilaton & $c_g ,  c_W , c_Z , c_\gamma ,c_{h^2}$
\\
Vectorlike Quarks &   $ c_g , c_\gamma , c_{Z\gamma} , c_{h^2} $ 
\\
Vectorlike Leptons &  $c_\gamma , c_{Z\gamma}, c_{h^2} $  
\\
New Charged Vector bosons &   $c_\gamma , c_{h^2} $ 
\\
Extra charged scalar bosons &  $c_g , c_\gamma , c_{Z\gamma}, c_{h^2}$ 
\\
\hline \hline
\end{tabular}
\caption{Nonvanishing $c_F$'s in various BSM's with an extra singlet scalar boson.  }
\label{tab:cici} 
\end{center}
\end{table}

\section{Fitting procedure} \label{sec:fitting}

\subsection{General parameterization and data}

Following the LHC Higgs working group \cite{LHCHiggsCrossSectionWorkingGroup:2012nn}, 
we find it useful to introduce model-independent notations to describe signal strength data: 
\beq
\kappa_i^2 \= \frac{\Gamma(h \to i)}{\Gamma(h\to i)_{SM}}, \qquad \kappa_H^2 \= \frac{ \Gamma^{tot}}{\Gamma^{tot}_{SM}}.
\label{eq:kappa1} \eeq
What is actually measured at LHC then can be written as
\beq
R^i_j \= R\left( \sigma(i \to h ) \frac{\Gamma(h\to j)}{\Gamma^{tot}} \right) \= \frac{ \kappa_i^2 \kappa_j^2}{ \kappa_H^2} \, \equiv \,  \hat{\kappa}_i^2 \hat{\kappa}_j^2, \qquad \mu_j \, \equiv \, \sum_i R^i_j 
\label{eq:kappa2} \eeq
where we define $\hat{\kappa}_i^2 = \kappa_i^2 / \kappa_H$. 
Signal strength in the final state $j$ for Higgs production modes combined is denoted by $\mu_j$. 
We discuss how our theory parameterization and the $\kappa_i$  notation are related in the later section.

As introduced in previous section, we parameterize effective couplings of Higgs-like boson by constants 
$b_i$, $c_i$ and the mixing angle $\alpha$ (and possible non-standard decay modes introduced below). 
We emphasize once again that these parameters do not multiplicatively parameterize signal strengths. 
For final states to which the Higgs boson couples at tree-level (such as $WW,ZZ$ and fermion pairs),
\beq 
\kappa_i^2 \= \frac{\Gamma(h \to i)}{\Gamma(h\to i)_{SM}} \= (b_i c_\alpha - c_i s_\alpha )^2 \qquad \textrm{for } i \= WW,\, ZZ,\, f\bar{f}
\eeq
where we denote $\cos \alpha = c_\alpha$ and similarly $\sin \alpha = s_\alpha$. For $gg$, $\gamma \gamma$ and $\gamma Z$ final states whose couplings to the Higgs boson are loop-induced,
\beq
\kappa_g^2 \= \frac{\Gamma(h \to gg)}{\Gamma(h\to gg)_{SM}} \=  (b_g c_\alpha - c_g s_\alpha )^2 \= \left( \, c_\alpha ( b_t {\cal C}_t \+ \Delta b_g ) \- c_g s_\alpha \, \right)^2
\ceq
\kappa_\gamma^2 \= \frac{\Gamma(h \to \gamma \gamma)}{\Gamma(h \to \gamma \gamma )_{SM}} \= (b_\gamma c_\alpha - c_\gamma s_\alpha )^2 \= \left( \, c_\alpha ( b_W {\cal B}_W \+ b_t {\cal B}_t \+ \Delta b_\gamma ) \- c_\gamma s_\alpha \, \right)^2
\label{eq:modifloop}\eeq
where the relative loop-functions of $W$ boson and top quark for $m_h = 125$ GeV are given in 
\Eq{eq:relloop}. Modifications to loop-induced decay \Eq{eq:modifloop} have several contributions; 
(i) from scalar mixing denoted by $\alpha$,  (ii) one inherit from singlet scalar couplings denoted by 
$c_{g,\gamma}$, (iii) from modification of top and/or $W$ boson couplings in the loop denoted by $b_t$ 
and $b_W$ weighted by loop factors introduced above, and (iv) last from some new physics effects 
directly modifying Higgs interaction eigenstate coupling denoted by $\Delta b_{g,\gamma}$. 

We also allow Higgs may have non-standard decay modes (such as invisible decay or flavor violating 
decay mode) which we parameterize by branching ratio into these modes, $BR_{nonSM}$. 
Total width is then written as
\beq
\kappa_H^2 \= \frac{\Gamma^{tot}}{\Gamma^{tot}_{SM}} \= \frac{ \sum_{i \ni SM} \, \kappa_i^2 BR(h \to i)_{SM} }{ 1- BR_{nonSM}}.
\eeq
 
We use the most up-to-date CMS and ATLAS data for $h \to ZZ, WW, \gamma \gamma, \tau \tau, b\bar{b}$ 
which is tabulated in Table~\ref{tab:data}. We naively use the best-fit signal strengths for each channel obtained 
with each best-fit $m_h$; refer to official notes of ATLAS \cite{ATLAS:comb} and CMS \cite{CMS:comb} for more 
dedicated study of signal strengths obtained with the same $m_h$. For the $\gamma\gamma$ and $ZZ$ modes, 
both ATLAS and CMS have analyzed the contributions from different production modes, but we use the combined results of multiple production modes since the uncertainty on the contribution of different production modes is 
rather large at the moment and the full covariance matrices on the errors are not available. We assume that all 
data are from $gg$-fusion and vector-boson fusion(VBF) in proportion to their production rates (except for 
$b\bar{b}$ mode which is assumed to be purely from $W/Z$ associate production) 
\beq
R( \sigma(pp \to h) ) \= \kappa_g^2 {\cal A}_g \+ \kappa_W^2 {\cal A}_W \+  \kappa_Z^2 {\cal A}_Z
\label{eq:prodmix} \eeq
where weighting factors for $m_h=125$GeV are
\beq
{\cal A}_g \= \frac{\sigma(ggF) }{ (\sigma(ggF)+ \sigma(VBF)) } \simeq 0.925, \qquad  {\cal A}_W \+ {\cal A}_Z\= \frac{\sigma(VBF)}{(\sigma(ggF)+\sigma(VBF))} \simeq 0.075.
\eeq
7 TeV and 8 TeV production rates are weighted-summed in proportion to luminosities accumulated 
in data we use. Likewise, $Vh$ associate production proceeds via either $W$ or $Z$ boson,
\beq
R( \sigma(pp \to Vh) ) \= \kappa_W^2 {\cal A}_W^\prime \+ \kappa_Z^2 {\cal A}_Z^\prime
\label{eq:prodmix2} \eeq
where ${\cal A}_W^\prime \+ {\cal A}_Z^\prime =1$.
We numerically checked that small mixed-in of VBF denoted by ${\cal A}_W+{\cal A}_Z$ does not significantly affect fit results. In this work, we simply ignore ${\cal A}_W+{\cal A}_Z$ contributions although we present general formula keeping its dependence for future reference. We do not consider $Z\gamma$ data because it has large uncertainty so far. However, by considering EWPT, Higgs coupling to $Z\gamma$ may also be constrained 
\cite{Falkowski:2013dza}, and this mode has important potential to discriminate various Higgs imposters 
\cite{Low:2012rj}. We use $m_h\=125$GeV when necessary.

\begin{table}[t] \centering \begin{tabular}{c|c|c|c|c}
\hline \hline
   & channel & luminosity ($fb^{-1}$) & $\mu$ & ref. \\
\hline \hline
   & $\gamma \gamma$ & 24.7 & $0.78^{+0.28}_{-0.26}$ & \cite{CMS:data1} \\   
   & $ZZ$ & 24.7 & $0.91^{+0.30}_{-0.24}$ & \cite{CMS:data2} \\
CMS   & $WW$ & 24.7 & $0.76^{+0.21}_{-0.21}$ & \cite{CMS:data3} \\
   & $\tau \tau$ & 24.3 & $1.1^{+0.4}_{-0.4}$ & \cite{CMS:data4} \\
   & $b\bar{b}$ & 17 & $1.3^{+0.7}_{-0.6}$ & \cite{CMS:data5} \\
\hline
   & $\gamma \gamma$ & 25 & $1.65^{+0.35}_{-0.30}$ & \cite{ATLAS:data1} \\   
   & $ZZ$ & 25 & $1.7^{+0.50}_{-0.40}$ & \cite{ATLAS:data2} \\
ATLAS   & $WW$ & 25 & $1.01^{+0.31}_{-0.31}$ & \cite{ATLAS:data3} \\
   & $\tau \tau$ & 18 & $0.7^{+0.7}_{-0.7}$ & \cite{ATLAS:data4} \\
   & $b\bar{b}$ & 18 & $-0.4^{+1.06}_{-1.06}$ & \cite{ATLAS:data5} \\
\hline \hline
\end{tabular} \caption{Signal strength data we use. 7TeV and 8TeV results are combined, and all subcategories of each final states are combined. We use best-fit signal strengths for each channel obtained with each best-fit $m_h$.}
\label{tab:data} \end{table} 

In all, we tabulate general parameterization of Higgs signal strengths we use in terms of $\kappa$ parameters in Table \ref{tab:gen-sig}.

\begin{table}[t] \centering \begin{tabular}{c||c|c}
\hline \hline
 & $pp \to h \to \gamma \gamma,\, WW,\,ZZ,\, \tau \tau$ & $ pp \to Vh \to Vb\bar{b}$ \\
\hline
signal strength $\mu_i$ & $\left(\hat{\kappa}_g^2 {\cal A}_g + \hat{\kappa}_{W}^2 {\cal A}_W + \hat{\kappa}_{Z}^2 {\cal A}_Z \right) \, \hat{\kappa}_i^2$  & $(\hat{\kappa}_{W}^2 {\cal A}_W^\prime + \hat{\kappa}_Z^2 {\cal A}_Z^\prime ) \, \hat{\kappa}_b^2$   \\
\hline \hline
\end{tabular} \caption{General parameterization of signal strength data that we use in terms of model independent $\hat{\kappa}$ parameters introduced in \Eq{eq:kappa1} and \Eq{eq:kappa2}. ${\cal A}_i$ and ${\cal A}_i^\prime$ are defined in \Eq{eq:prodmix} and \Eq{eq:prodmix2}.}
\label{tab:gen-sig} \end{table}

\subsection{SM fit}

First of all, pure SM fit is performed. We use MINUIT package~\cite{MINUIT} to carry out $\chi^2$-fit:
\beq
\chi^2/\nu \= 12.01/10 \= 1.20 \quad \textrm{(both)}
\ceq
\chi^2/\nu \= 2.33/5 \= 0.466 \quad \textrm{(CMS)}
\ceq
\chi^2/\nu \= 9.69/5 \= 1.94 \quad \textrm{(ATLAS)}.
\eeq

\subsection{Preliminary: fits without extra singlet scalar} \label{sec:bi-fit}

In this subsection, we use only $b_i$ to carry out best-fit. When new particles directly couple to Higgs 
boson in a gauge invariant way under the SM gauge group and modify Higgs couplings, the model falls 
into this category. This study also allows us to compare our fit results with other results available 
in literature, and moreover to discuss similarities and differences between individual ATLAS and CMS data. 

We simply assume custodial symmetry for the SM Higgs couplings to the weak gauge bosons: 
\beq
b_W \= b_Z \, \equiv \, b_V.
\eeq
We also simply treat all fermions couplings universally
\beq
b_t \= b_b \= b_\tau \, \equiv \, b_f
\eeq
although we checked that allowing $b_t$ to float independently would not qualitatively modify our statements.

We consider following four cases: (results are also summarized in Table \ref{tab:fit-bi})
\bei
\item  $\{ \, \Delta b_\gamma \, \}$:
This may represent some models of extra leptons \cite{Moreau:2012da} or $W^\prime$ 
or extra charged scalar. In this case,
\beq
\kappa_\gamma^2 \= b_\gamma^2 \= (1+\Delta b_\gamma)^2, \quad \kappa_g^2 \= \kappa_{V}^2 \= \kappa_f^2 \= 1, \quad \kappa_H^2 \, \simeq \, 1
\eeq
giving best-fit
\beq
\Delta b_\gamma \= 0.090^{+0.0889}_{-0.0999}, \quad \chi^2/\nu \= 11.19/9\=1.24 \quad \textrm{(both)}
\ceq
\Delta b_\gamma \= -0.117^{+0.147}_{-0.162}, \quad \chi^2/\nu \= 1.71/4\=0.428 \quad \textrm{(CMS)}
\ceq
\Delta b_\gamma \= 0.28^{+0.134}_{-0.118}, \quad \chi^2/\nu \= 4.99/4\=1.25\quad \textrm{(ATLAS)}
\eeq
We show best-fit using both ATLAS and CMS data, as well as best-fit obtained from individual data. $\mu_\gamma \= 1.19$ (both).

\item $\{\, \Delta b_g , \, \Delta b_\gamma \,\} $:
This may represent some models of extra quarks \cite{Moreau:2012da}. In this case, 
\beq
\kappa_g^2 \= b_g^2 \= (1+\Delta b_g)^2, \quad \kappa_\gamma^2 \= b_\gamma^2 \= (1+\Delta b_\gamma)^2, \quad \kappa_{mix}^2 \= 1, 
\ceq
\kappa_H^2 \, \simeq \,\kappa_g^2 \, Br(h \to gg) \+ Br(h \to others)
\eeq
giving best-fit ($\mu_\gamma \= 1.19$ (both))
\beq
\Delta b_g \= -0.0180^{+0.0559}_{-0.0577}, \quad \Delta b_\gamma \= 0.107^{+0.0916}_{-0.100}, \quad \chi^2/\nu \= 11.13/8\=1.39 \quad \textrm{(both)}
\ceq
\Delta b_g \= -0.078^{+0.0760}_{-0.0784}, \quad \Delta b_\gamma \= -0.048^{+0.157}_{-0.175}, \quad \chi^2/\nu \= 0.859/3\=0.286 \quad \textrm{(CMS)}
\ceq
\Delta b_g \= 0.11^{+0.0867}_{-0.0830}, \quad \Delta b_\gamma \= 0.17^{+0.117}_{-0.113}, \quad \chi^2/\nu \= 4.14/3\=1.38\quad \textrm{(ATLAS)}
\eeq

\item  $\{ \, b_V, \, b_f \, \}$:
This may represent some simple composite Higgs models, see e.g. Ref.\cite{Azatov:2012bz, Espinosa:2012im}. In this case,
\beq
\kappa_g^2 \= b_f^2, \quad \kappa_\gamma^2 \= ( b_V {\cal B}_W \+ b_f {\cal B}_t )^2, \quad \kappa_V^2 \= b_V^2, \quad \kappa_f^2 \= b_f^2
\ceq
\kappa_H^2 \= \sum_i \kappa_i^2 \, BR(i)_{SM}.
\eeq
giving best-fit
\beq
b_V \= 1.031^{+0.0682}_{-0.0688}, \quad b_f \= 0.962^{+0.124}_{-0.124}, \quad \chi^2/\nu \= 11.74/8 \= 1.47 \quad \textrm{(both)}
\ceq
b_V \= 0.898^{+ 0.081}_{-0.081}, \quad b_f \= 1.021^{+0.137}_{-0.154}, \quad \chi^2/\nu \= 0.808/3 \= 0.27 \quad \textrm{(CMS)}
\ceq
b_V \= 1.345^{+0.162}_{-0.144}, \quad b_f \= 0.808^{+0.144}_{-0.117}, \quad \chi^2/\nu \= 4.52/3 \= 1.51 \quad \textrm{(ATLAS)}
\eeq

\item  $\{ \, b_V, \, b_u, \, b_d, \, (c_\alpha)\, \}$:
We relax assumptions of universal fermion couplings, and allow all up-type quark couplings are 
modified by $b_u$, and all up-type quarks and charged leptons are modified by $b_d$. 
We further assume that
\beq
b_V \leq 1 .
\eeq
This ansatz can represent various two-Higgs doublet model when both of two Higgs doublets develop nonzero
VEV's and contribute to the EWSB. When $c_\alpha$ is allowed, this ansatz may further consider 
as a doublet+singlet scenario such as the scalar sector of next-to-MSSM (NMSSM). Generally,
\beq
\kappa_g^2 \= b_u^2 c_\alpha^2, \quad \kappa_\gamma^2 \= ( b_V {\cal B}_W \+ b_u {\cal B}_t )^2 c_\alpha^2, \quad \kappa_V^2 \= b_V^2 c_\alpha^2, \quad \kappa_u^2 \= b_u^2 c_\alpha^2, \quad \kappa_d^2 \= b_d^2 c_\alpha^2
\ceq
\kappa_H^2 \= \sum_i \kappa_i^2 \, BR(i)_{SM}.
\eeq
With $c_\alpha=1$ fixed, best-fit is
\beq
b_V \= 1.0_{-0.0601}, \quad b_u \= 0.969^{+0.0632}_{-0.0647}, \quad b_d \= 0.938^{+0.0899}_{-0.0788}, 
\ceq
\chi^2/\nu \= 11.86/7 \= 1.69 \quad \textrm{(both)}.
\eeq
Interestingly, $b_V=1$ gives the best-fit. Another interesting result is that allowing $c_\alpha$ doesn't change best-fit results; $c_\alpha=1$ gives (the same) best-fit. For reference, if we allowed $b_V$ to vary above $1$, we would have obtained $b_V=1.05$ as a best-fit.

\item $\{\, \Delta b_g, \, \Delta b_\gamma, \, b_V, \, b_f \,\}$:
In this case,
\beq
\kappa_g^2 \= b_g^2 \= (b_f +\Delta b_g)^2, \quad \kappa_\gamma^2 \= b_\gamma^2 \= (b_V {\cal B}_W \+ b_f {\cal B}_t + \Delta b_\gamma)^2 , \quad \kappa_V^2 \= b_V^2, \quad \kappa_f^2 \= b_f^2
\ceq
\kappa_H^2 \= \sum_i \kappa_i^2 \, BR(i)_{SM}.
\eeq
giving best-fit
\beq
\Delta b_g \= 0.041^{+0.0596}_{-0.0621}, \quad \Delta b_\gamma \= 0.117^{+0.0927}_{-0.101}, \quad b_V \= 0.941^{+0.0569}_{-0.0581}, \quad b_f \=0.961^{+0.116}_{-0.127},
\ceq
\chi^2 /\nu \= 11.07/6 \= 1.85.
\eeq
We do not consider fitting to individual ATLAS and CMS data here because there are too small number of degrees of freedom $(\nu=1)$ which may not allow meaningful statistical interpretation of fit results.

\eei

\begin{table}[t] \centering \begin{tabular}{c||c|c}
\hline \hline
                & our fits &  fits in other refs. \\
\hline \hline
$(\, \Delta b_g, \, \Delta b_\gamma \,)$ & $(\, -0.0180^{+0.0559}_{-0.0577}, \, 0.107^{+0.0916}_{-0.100} \,)$ & $(\, -0.12^{\pm 0.11},\, 0.18^{\pm0.12} \,)$ \cite{Ellis:2013lra}\\
 &   & $(\, -0.083^{\pm 0.067},\, 0.13^{\pm0.12} \,)$ \cite{Falkowski:2013dza}\\
 &   & Fig.5 of Ref.\cite{Giardino:2013bma} \\
(ATLAS-only) & $(0.11^{+0.0867}_{-0.0830}, \, 0.17^{+0.117}_{-0.113})$ & $(0.08^{\pm0.14},\, 0.23^{+0.16}_{-0.13})$ \cite{ATLAS:comb} \\
\hline
$(\, b_V,\, b_f \,)$ & $(\, 1.031^{+0.0682}_{-0.0688}, \, 0.962^{+0.124}_{-0.124}  \,)$ & $(\, 1.03^{\pm 0.06},\, 0.84^{\pm0.15} \,)$ \cite{Ellis:2013lra}\\
 &   & Fig.3 of Ref.\cite{Falkowski:2013dza}, Fig.4 of Ref.\cite{Giardino:2013bma}\\
 (ATLAS-only) & $(1.345^{+0.162}_{-0.144}, \, 0.808^{+0.144}_{-0.117})$ & $(1.13^{\pm0.08},\, 0.90^{\pm0.17})$ \cite{ATLAS:comb} \\
\hline \hline
\end{tabular} \caption{Comparison of our fit results with results available in other literature. Only results based on up-to-date data after Moriond 2013 are compared. We sometimes re-interpret other's results in accordance with our notation. If only best-fit figure is available, we cite relevant figure and reference. Cases that are not shown here do not have equivalent results in literature.}
\label{tab:compare-others} \end{table}

\begin{table}[t] \centering \begin{tabular}{c||c|c|c}
\hline \hline
                & both & CMS & ATLAS \\
\hline \hline
SM         & $\chi^2/\nu = 12.01/10 = 1.20$  &  $2.33/5 =0.466$  &  $9.69/5 = 1.94$ \\
\hline \hline
$(\, \Delta b_\gamma \,)$ & (0.090) & (-0.117) & (0.28)  \\
 & 11.19/9=1.24 & 1.71/4=0.428 & 4.99/4=1.25  \\
\hline
$(\, \Delta b_g, \Delta b_\gamma \,)$ & (-0.018, 0.107) & (-0.078, -0.048) & (0.11, 0.17) \\
 & 11.13/8 = 1.39 & 0.859/3 = 0.286 & 4.14/3 = 1.38 \\
\hline
 $(\, b_V,\, b_f \,)$ & $(\, 1.031, \, 0.962 \,)$ & $(\, 0.898,\, 1.021 \,)$ & $(\, 1.345, \, 0.808 \,)$ \\
  & $11.74/8 = 1.47 $ & 0.808/3=0.27 & 4.52/3=1.51 \\
\hline
 $(\, b_V\leq1,\, b_u,\,b_d \,)$ & $(\, 1.0, \, 0.969,\, 0.938 \,)$ &  & \\
  & $11.86/7 = 1.69 $ &  &  \\
\hline
 $(\, \Delta b_g, \, \Delta b_\gamma, \, b_V, \, b_f \,)$ & $(\, 0.041,\, 0.117, \qquad$ & & \\
  & $\qquad \qquad 0.941,\, 0.961 \,)$ & &  \\
  & 11.07/6 = 1.85 &  &  \\
\hline \hline
\end{tabular} \caption{Best-fit results using $b_i$ only from both CMS and ATLAS data as well as individual. Errors are shown in text.}
\label{tab:fit-bi} \end{table}

We compare our fit results with other results available in literature. For proper comparison, we use other results based on up-to-date data after Morion 2013. As tabulated in Table \ref{tab:compare-others}, we obtain 
fairly good agreement on central values and sizes of uncertainties. Some difference may originate from different 
datasets used; we are using only inclusively combined data while most other literatures uses more individual 
production modes. Another important difference is that we naively use best-fit signal strengths each fitted with 
different best-fit $m_h$ values while ATLAS official document \cite{ATLAS:comb} uses signal strengths obtained 
all with the same $m_h=125.5$ GeV. For example, our $\mu_{ZZ}=1.7$ is obtained with best-fit 
$m_h=124.3$ GeV while ATLAS uses $\mu_{ZZ}=1.5$ \cite{ATLAS:comb} obtained with assumed 
$m_h=125.5$ GeV. This causes some difference in best-fit $b_V$ values. We note that official study of 
coupling fit from CMS \cite{CMS:comb} is not updated with up-to-date data yet. 

From cases (1), (2) and  (3) that are also tabulated in Table~\ref{tab:fit-bi}, one can also learn about general 
trends of present ATLAS and CMS data. Generally, CMS data has better fit to $\{ b_i \}$ parameterization than 
ATLAS data does. CMS data generally prefers the suppression of signal strengths while ATLAS data prefers 
the enhancement. However, each best-fit parameters are marginally consistent with each other (between 
CMS-only and ATLAS-only fit results) within combined $2\sigma$ uncertainties although $b_V$ best-fit results 
are a bit more discrepant. Consequently, best-fits in terms of $\{ b_i \}$ to both ATLAS and CMS data simultaneously is not improved significantly 
from pure SM fit. This may or may not mean that $\{ b_i \}$ 
parameterization is disfavored partly because fits to individual data are indeed improved.

\section{Theories of an extra singlet scalar mixed in} \label{sec:sing-fit}

We now consider an extra singlet scalar mixing with SM Higgs. In the remainder of the section, we make following simplifying but general enough assumptions:
\bei

\item Singlet has only loop-induced couplings to SM particles; only $c_g,\, c_\gamma$ are non-zero among all $c_i$.  
\item However, singlet may have non-standard decay modes: $BR_{nonSM} \geq 0$. This parameter also considers a possibility that $m_{H_2} \leq m_{H_1}/2=62.5$GeV.
\item Higgs couplings are not directly modified: all $b_i = 1,\, \Delta b_i=0$. But we discuss briefly the case with non-zero $\Delta b_i$ in Sec.\ref{sec:abc-fit}.

\eei
With these assumptions, we have following three model-independent parameters that can parameterize signal strength data
\beq
\{ \, \hat{\kappa}_g, \quad \hat{\kappa}_{\gamma}, \quad \hat{\kappa}_{mix} \, \}
\eeq 
where $\kappa_{mix}$ globally parameterizes all SM decay widths other than $gg$ and $\gamma \gamma$ modes defined as 
\beq
\kappa_{mix}^2 \= \frac{\Gamma(H_1 \to WW)}{\Gamma(h\to WW)_{SM}}.
\eeq
Again, $\hat{\kappa}_i^2 = \kappa_i^2 / \kappa_H$. See Table~\ref{tab:fit1} for how $\kappa$s contribute to signal strengths, where different production modes are weighted-summed with production ratio as discussed previously. This parameterization has not been considered yet by the LHC Higgs working group \cite{LHCHiggsCrossSectionWorkingGroup:2012nn}, and this type of model has only been briefly 
studied in Refs. \cite{Carmi:2012yp, Carmi:2012in}. 

\begin{table}[t] \centering \begin{tabular}{c||c|c|c}
\hline \hline
 & $pp \to h \to \gamma \gamma$ & $ pp \to h \to WW,\,ZZ,\,  \tau \tau$ & $ pp \to Vh \to Vb\bar{b}$ \\
\hline
$\mu_i \, $ & $\left(\hat{\kappa}_g^2 {\cal A}_g + \hat{\kappa}_{mix}^2 ({\cal A}_W+{\cal A}_Z) \right) \hat{\kappa}_\gamma^2$ & $\left(\hat{\kappa}_g^2 {\cal A}_g + \hat{\kappa}_{mix}^2 ({\cal A}_W+{\cal A}_Z) \right) \hat{\kappa}_{mix}^2$ & $\hat{\kappa}_{mix}^2 \hat{\kappa}_{mix}^2$   \\
\hline \hline
\end{tabular} \caption{Parameterization of signal strengths in extra singlet models with our assumptions. ${\cal A}_{g,W,Z}$ are defined in \Eq{eq:prodmix}.}
\label{tab:fit1} \end{table} 

In this section, we consider several special cases of this general parameterization. All best-fit results are summarized in Table \ref{tab:c-fit}.

\subsection{How likely are signals universally modified? : $\{ \, \alpha, \, BR_{nonSM} \, \}$}
We begin by evaluating how likely that all Higgs signal strengths are universally modified. 
Universal modification can arise through singlet portal mixing and/or non-SM decay width. 

In this case, general parameterization introduced above are simplified, and one common parameter
\beq
\hat{\kappa}^2_{univ} \, \equiv \, \frac{ \kappa^2_{univ} }{ \kappa_H }
\eeq
universally parameterizes all Higgs signal strengths as
\beq
\mu_i \= \kappa^2_{univ} \frac{\kappa^2_{univ}}{\kappa_H^2} \= \hat{\kappa}_{univ}^2 \hat{\kappa}_{univ}^2.
\eeq
This parameterization shows that it is difficult to know whether the universal modification of signal strengths is originated from $\kappa_{univ}$(singlet portal mixing) or $\kappa_H$(non-SM decay mode). Best-fit is
\beq
\hat{\kappa}^2_{univ} \= 1.012^{+0.0517}_{-0.0549}, \quad  \chi^2/\nu \= \frac{11.96}{9} \= 1.33 \qquad \textrm{(both)}
\label{eq:fit-univ}
\ceq
\hat{\kappa}^2_{univ} \= 0.930^{+0.0675}_{-0.0710}, \quad  \chi^2/\nu \= \frac{1.25}{4} \= 0.31 \qquad \textrm{(CMS)}
\ceq
\hat{\kappa}^2_{univ} \= 1.149^{+0.0762}_{-0.0812}, \quad  \chi^2/\nu \= \frac{6.53}{4} \= 1.63 \qquad \textrm{(ATLAS)}
\eeq
$1,2\sigma$ ranges of best-fit parameters are defined as ranges for $\Delta \chi^2 =1,4$ from a single variable cumulative distribution function of $\chi^2$ assuming no correlations, where $\chi^2 = \chi^2_{min} + \Delta \chi^2$. Again, fit is not improved from pure SM fit if both data are used while fits to individual data are improved. Almost no universal modification is preferred. This is partially resulted from the fact that CMS prefers suppression while ATLAS prefers enhancement. 

\begin{figure} \centering
\includegraphics[width=0.45\textwidth]{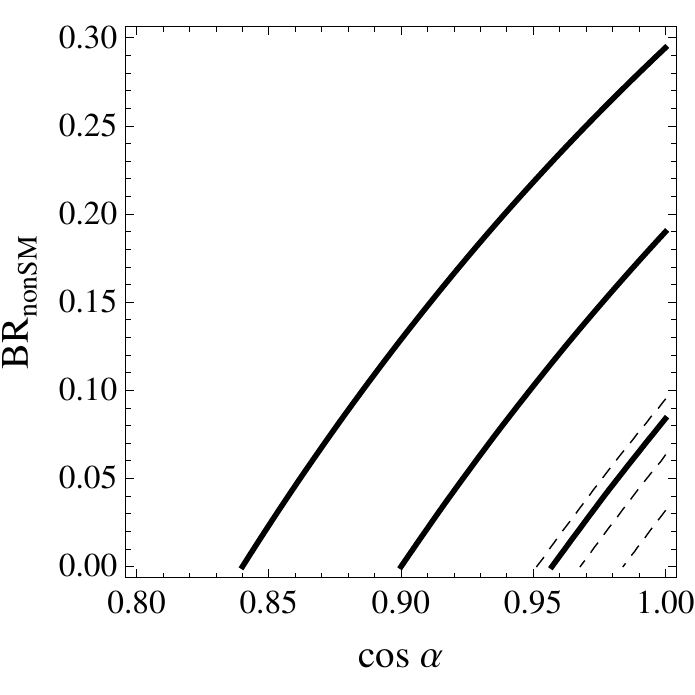}
\caption{$1, 2, 3 \sigma$ ranges of best-fit is shown for the case of universal modification. Best-fit is given by \Eq{eq:fit-univ} as well as \Eq{eq:univ-brnon} and \Eq{eq:univ-mix}. Dashed lines are expected if all future data are $R^i_j = 1.0 \pm 0.1$.}
\label{fig:contour-univ}
\end{figure}

In the end,  there are two theory parameters $\{ \, \alpha, \, BR_{nonSM} \, \}$ for the universal modification of 
Higgs signal strengths, with the following relations: 
\beq
\kappa_{univ}^2 \= c_\alpha^2, \qquad \kappa_H^2 \= \frac{c_\alpha^2}{1-BR_{nonSM}}.
\eeq
The 2D contours of $\Delta \chi^2$ in the theory parameters plane is shown in \Fig{fig:contour-univ}. 
If either $c_\alpha =1 $ or $BR_{nonSM} =0$ is fixed, 
\bea
BR_{nonSM} &=& -0.0241^{+0.108}_{-0.107} \qquad \textrm{for } c_\alpha\=1 \label{eq:univ-brnon}\\
c_\alpha &=& 1.012^{+0.0517}_{-0.0549} \qquad \textrm{for } BR_{nonSM}\=0.  \label{eq:univ-mix}
\eea
Thus, $BR_{nonSM} \leq 18.8\%$ at 95\%C.L. if $c_\alpha \= 1$ fixed as also similarly obtained by Ref.\cite{Falkowski:2013dza,Giardino:2013bma,Ellis:2013lra} using up-to-date data. Likewise, $c_\alpha \geq 0.904$ at 95\%C.L. if $BR_{nonSM}=0$ fixed.

Suppose future data is $R^i_j = 1.0 \pm 0.2 \,(0.1)$ for all 10 channels we are considering. 
This would be a perfect case for the SM Higgs boson.  The best-fit then would yield
\beq
\hat{\kappa}_{univ}^2 \= 1.0^{\+ 0.0311 \,(0.0157)}_{\- 0.0321\,(0.0159)}.
\eeq
Corresponding 2D $\Delta \chi^2$ contours are also shown in dashed curves in 
\Fig{fig:contour-univ} to illustrate how much we are improved, and how much we are still not able 
to probe. The best fit would imply $BR_{nonSM} \leq 12.4 \,(6.2)\%$ at 95\%C.L. if $c_\alpha =1$ fixed, 
or $c_\alpha \geq 0.94\,(0.97)$ if $BR_{nonSM}=0$ fixed. This discussion may help us grasp how well 
one can constrain universal-suppression models with future data.

\subsection{Models with extra leptons or $W^\prime$ : $\{\, \alpha,\, c_\gamma \,\}$} \label{sec:exlep}

\begin{figure} \centering
\includegraphics[width=0.47\textwidth]{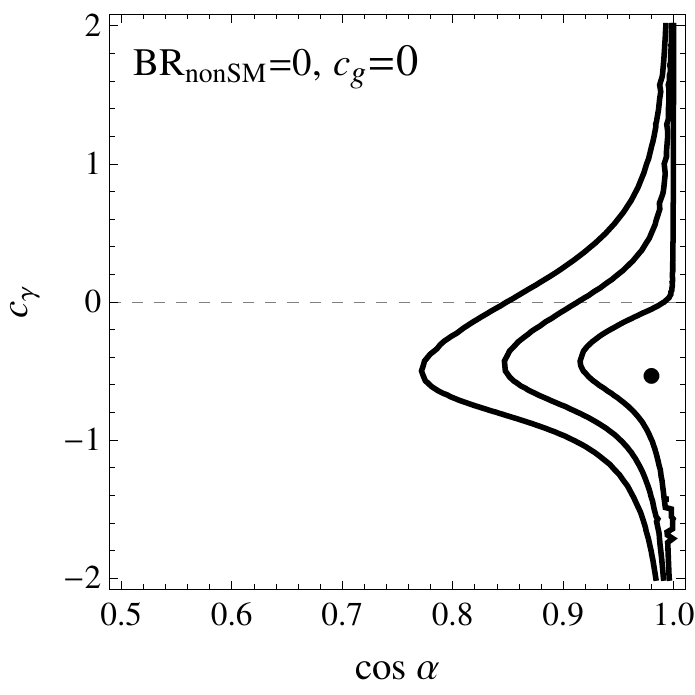}
\caption{ $1, 2, 3 \, \sigma$ ranges of best-fit for $\{ \alpha, \, c_\gamma \}$.}
\label{fig:contour-ca-cr}
\end{figure}

Extra leptons or $W^\prime$ induce singlet couplings to photons at one-loop. 
The free parameters to fit are  $\{\, \alpha,\, c_\gamma \,\}$.    In this case,
\beq
\kappa_\gamma^2 \= (c_\alpha - c_\gamma s_\alpha)^2, \quad \kappa_g^2 \= \kappa_{mix}^2 \= 
c_\alpha^2, \quad \kappa_H^2 \, \simeq \, c_\alpha^2
\eeq
where we ignored small diphoton decay modes in total width. Therefore, two parameters $\hat{\kappa}_\gamma$ and $\hat{\kappa}_{mix}=\hat{\kappa}_g$ can parameterize all signal strengths. Best-fit is
\beq
c_\alpha \= 0.98_{-0.056}, \quad c_\gamma \= -0.55^{+0.50}_{-0.45}, \quad \chi^2/\nu \= 11.1/8 \=1.39
\label{eq:fit-alcgam} \eeq
corresponding to $\mu_\gamma = 1.19$. Preferred parameter space is shown in \Fig{fig:contour-ca-cr} left panel. Although total $\chi^2$ of the fit is improved from SM fit, $\chi^2/\nu$ is not.

\begin{figure} \centering
\includegraphics[width=0.95\textwidth]{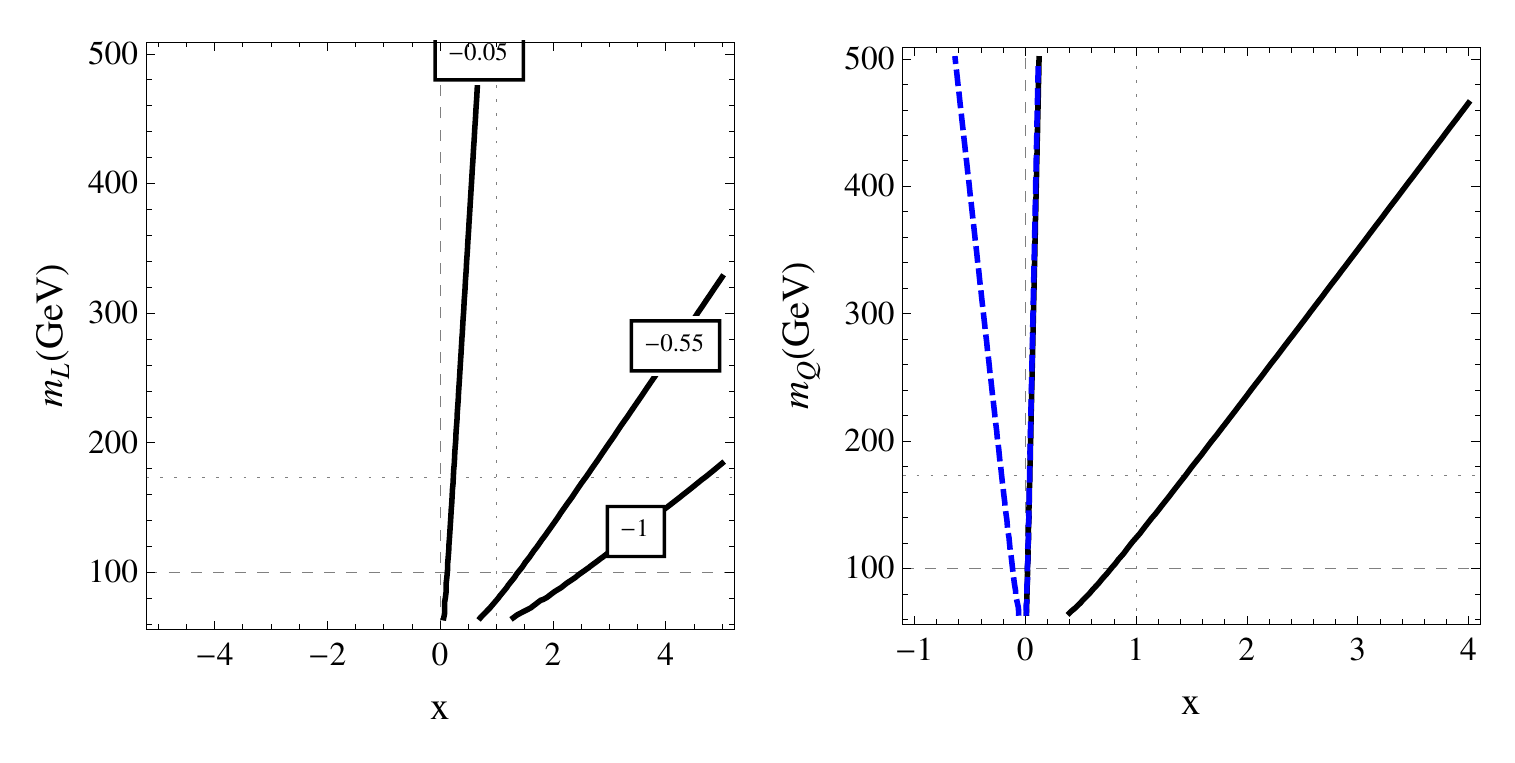}
\caption{$\pm1 \sigma$ favored region from best-fits of $c_\gamma$ (black solid) and/or $c_g$ (blue dashed) in models of (Left): vector-like lepton considered in case(2), and (Right): top-like quark considered in case(3). }
\label{fig:vec-lepqua}
\end{figure}

We attempt to interpret the best-fit results in terms of specific underlying models. 
The example model \cite{Batell:2012zw} contains a singlet scalar which couples to two sets of vector-like 
fermions carrying electric charge 1. The $c_\gamma$ can be written as a function of the lightest fermion mass, 
$m_L$, and its yukawa coupling to singlet, $x$, as 
\beq
c_\gamma \=  \frac{v}{\sqrt{2}A^{SM}_\gamma} \, \frac{xA_{1/2}(m_{H_1}^2/4m_L^2)}{m_L}.
\eeq
Since $c_\gamma$ is fitted from 125GeV data, $m_{H_1}=125$GeV. The favored parameter space of $m_L$ and $x$ 
is plotted in \Fig{fig:vec-lepqua} left panel. As SM is consistent with fit and errors are 
somewhat large, we do not obtain strong constraints on this model. However, one can already observe that 
the majority of parameter space favored within $\pm1 \sigma$ range involve yukawa couplings stronger 
than top yukawa and/or fermions lighter than top quarks. 

\subsection{Models with extra quarks : $\{ \, \alpha,\, c_\gamma,\, c_g \, \}$}

\begin{figure} \centering
\includegraphics[width=0.95\textwidth]{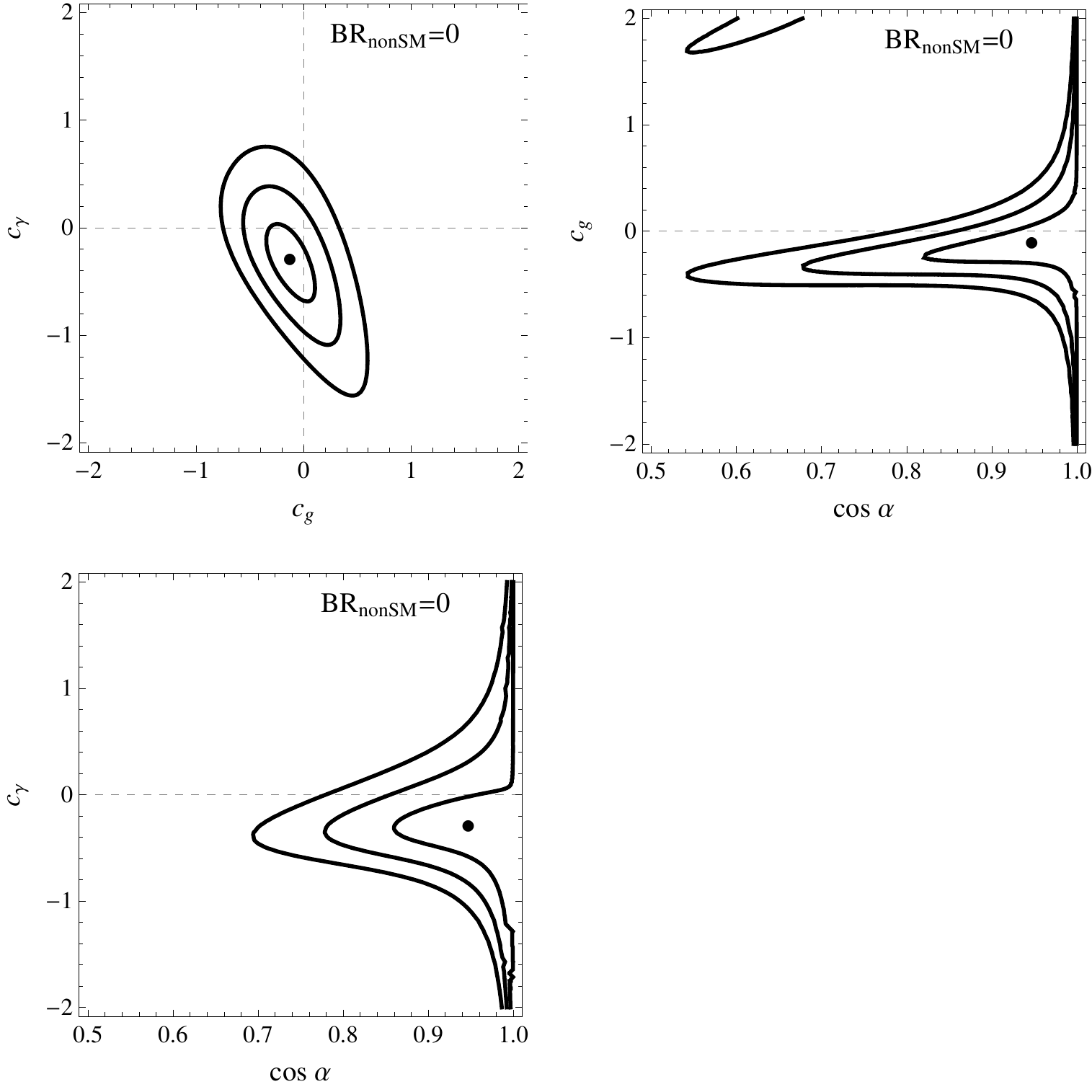}
\caption{$1, 2, 3 \sigma$ regions around the best-fit in the case of $\{ \alpha, c_\gamma, c_g \}$ without non-standard decay modes.}
\label{fig:contour2-thy}
\end{figure}

Extra quarks can induce singlet couplings to photons as well as gluons at one-loop. 
The free parameters to fit are  $\{ \, \alpha,\, c_\gamma,\, c_g \, \}$, and we have 
\beq
\kappa_\gamma^2 \= (c_\alpha - c_\gamma s_\alpha)^2, \quad \kappa_g^2 \= (c_\alpha - c_g s_\alpha)^2, \quad \kappa_{mix}^2 \= c_\alpha^2, \quad \kappa_H^2 \, \simeq \,  0.0857\, \kappa_g^2 \+ 0.9143 \, c_\alpha^2 
\eeq
Therefore, three parameters $\hat{\kappa}_g$, $\hat{\kappa}_\gamma$ and $\hat{\kappa}_{mix}$ can parameterize all signal strengths. Best-fit is
\beq
c_g \= -0.128^{+0.185}_{-0.174}, \quad c_\gamma \= -0.313^{+0.296}_{-0.269}, \quad c_\alpha \= 0.947_{-0.0873}, \quad \chi^2/\nu \= 11.1/7 \= 1.58
\label{eq:bestfit2-thy} \eeq
and various 2D contours of $\Delta \chi^2$ are shown in \Fig{fig:contour2-thy}. Compared to extra lepton models, total $\chi^2$ of the best fit is not improved, whereas somewhat larger mixing is preferred.  

We take a simple toy model involving a singlet scalar coupling to a single extra quark whose quantum 
numbers are identical to top quarks. The quarks generate both $c_\gamma$ and $c_g$. 
With the lightest extra quark mass $m_Q$ and its coupling $x$ to the singlet scalar, loop-induced couplings are 
\beq
c_\gamma \= \frac{v}{A_\gamma^{SM}} \, \frac{ x N_cQ_t^2 A_{1/2}(m_{H_1}^2/4m_Q^2)}{m_Q}
\ceq
c_g \= \frac{ x v A_{1/2}(m_{H_1}^2/4m_Q^2)}{A_{1/2}(m_{H_1}^2/4m_t^2) m_Q}
\eeq
where $m_{H_1}=125$GeV. In this simple model, the parameter space favored by $c_\gamma$ and $c_g$ are separately shown in \Fig{fig:vec-lepqua} right panel; two regions barely overlap. The best-fit solution of $c_g$ prefers to negative sign of yukawa coupling $x$ while $c_\gamma$ prefer to positive sign at $1\sigma$ level.

\subsection{Non-standard decay modes and an upper bound : 
$\{ \, \alpha,\, c_\gamma,\, c_g \, BR_{nonSM} \, \}$}
As briefly discussed in universal modification case, if both $\alpha$ and $BR_{nonSM}$ are involved, no unique solution of best fit is found. Let us elaborate this difficulty, how to handle this difficulty, and how we generically obtain upper bound on $BR_{nonSM}$ in the singlet extended models. We first consider following most general case based on our assumptions: $\{ \, \alpha,\, c_\gamma,\, c_g \, BR_{nonSM} \, \}$, with 
\beq
\kappa_g^2 \= (c_\alpha - c_g s_\alpha)^2, \quad \kappa_\gamma^2 \= ( c_\alpha - c_\gamma s_\alpha)^2, \quad\kappa_{mix}^2 \= c_\alpha^2, \quad \kappa_H^2 = \frac{0.0857 \kappa_g^2 \+ 0.9143 \kappa_{mix}^2}{1 \- BR_{nonSM} }.
\label{eq:nonsmwid} 
\eeq
Therefore, three parameters $\hat{\kappa}_g$, $\hat{\kappa}_\gamma$ and $\hat{\kappa}_{mix}$ can parameterize all signal strengths. We do not obtain unique solution of best-fit because theory has four free parameters while data is parameterized by three parameters. Alternatively, one can note that $\hat{\kappa}$s are invariant under the following transformation of theory parameters $c \to c^\prime$
\beq
\frac{ c_\alpha^{\prime2}}{c_H^\prime} \= \frac{c_\alpha^2}{c_H}, \qquad c_g^\prime \= c_g \frac{s_\alpha}{c_\alpha} \frac{c_\alpha^\prime}{s_\alpha^\prime}, \qquad c_\gamma^\prime \= c_\gamma \frac{s_\alpha}{c_\alpha} \frac{c_\alpha^\prime}{s_\alpha^\prime} \qquad \To \qquad \hat{\kappa}^\prime \= \hat{\kappa}
\label{eq:c-transf}\eeq
which explains redundancy in this four-parameter description of data.

We thus rather fit three hatted theory parameters 
\beq
\{ \,\hat{c}_g, \quad \hat{c}_\gamma, \quad \hat{c}_\alpha \, \}
\eeq
 defined as
\beq
\hat{\kappa}_{mix}^2 \= \frac{c_\alpha^2}{\kappa_H} \, \equiv \, \hat{c}_\alpha^2, \quad \hat{\kappa}_g^2 \= \frac{( c_\alpha - c_g s_\alpha)^2}{\kappa_H} \, \equiv \, (\hat{c}_\alpha - \hat{c}_g \hat{s}_\alpha)^2,  \quad \hat{\kappa}_\gamma^2 \= \frac{(c_\alpha - c_\gamma s_\alpha)^2}{\kappa_H} \, \equiv \, (\hat{c}_\alpha - \hat{c}_\gamma \hat{s}_\alpha)^2. 
\eeq
Hatted cosine and sine are defined to obey $\hat{s}_\alpha^2 + \hat{c}_\alpha^2 \= 1$ with a reasonable assumption $\hat{c}_\alpha^2 \leq 1$. This assumption however does not affect the result of best-fit. Solving these in terms of un-hatted parameters,
\beq
c_\alpha^2  \= \kappa_H \hat{c}_\alpha^2, \quad c_g \= \hat{c}_g \frac{ \hat{s}_\alpha}{ \hat{c}_\alpha} \frac{c_\alpha}{s_\alpha} \= \hat{c}_g \frac{ \hat{s}_\alpha}{ \hat{c}_\alpha} \frac{\sqrt{\kappa_H} \hat{c}_\alpha}{\sqrt{1-\kappa_H \hat{c}_\alpha}}, \quad c_\gamma \= \hat{c}_\gamma \frac{ \hat{s}_\alpha}{ \hat{c}_\alpha} \frac{c_\alpha}{s_\alpha}
\label{eq:c-traj} \eeq 
which is nothing but the transformation \Eq{eq:c-transf} between $c_i$ and $\hat{c}_i$ assuming that $\hat{c}_i$ are theory parameters $c_i$ with $\kappa_H=1$.

\begin{figure} \centering
\includegraphics[width=0.55\textwidth]{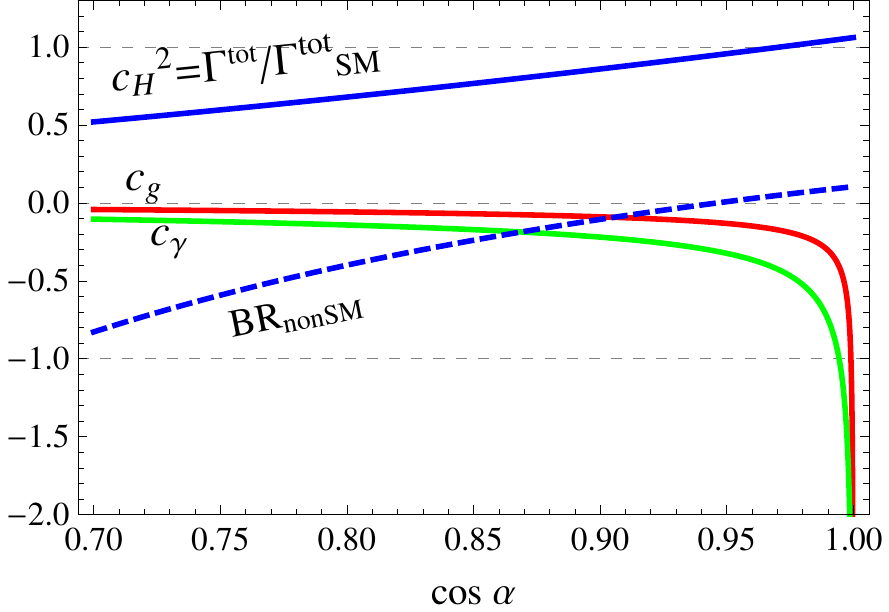}
\caption{Trajectories of theory parameters giving the best-fit in \Eq{eq:bestfit-thy}. Any choice of $\cos \alpha$ and corresponding parameters on the trajectories give equivalently good fits to data. Curves are obtained by the transformation \Eq{eq:c-traj}. }
\label{fig:phys-cos}
\end{figure}

Best-fit in terms of hatted parameters is
\beq
\hat{c}_g \= -0.176^{+0.231}_{-0.219}, \quad \hat{c}_\gamma \= -0.432^{+0.406}_{-0.374}, \quad \hat{c}_\alpha \= 0.971_{-0.0451}, \quad \chi^2/\nu \= 11.1/7 \= 1.58.
\label{eq:bestfit-thy} \eeq
The previous case (3) with $\{ \alpha, c_\gamma, c_g \}$ is one special point of four-parameter fit here. This can be seen from the trajectories of theory parameters in \Fig{fig:phys-cos} giving equivalently good fits to data; $BR_{nonSM}$ vanishes at previous best-fit $c_\alpha=0.947$ and other parameters there correspond to best-fit values of case (3). 

If mixing is too large $(c_\alpha \lesssim 0.947)$, the central value of best-fit non-SM width becomes negative. If one requires the positivity of non-SM width, this large mixing region would have poorer fit. On the other hand, if mixing is too small, then required $c_{g,\gamma}$ become very large. This might indicate either charged particles lighter than $W$ boson or coupling much stronger than electroweak coupling, thus may not be favored or calculable.

Now we see how Higgs total decay width or non-standard branching ratio can be upper bounded from the fact that $c_\alpha^2 \leq 1$. If $\kappa_H$ is too large, even maximally possible $c_\alpha^2 \=1 $ cannot make $\hat{c}_\alpha^2$ large enough to satisfy the lower-bound of best-fit result. Using the relation $\kappa_H=c_\alpha^2/\hat{c}_\alpha^2$, this argument leads to the following upper bound on $\kappa_H$:
\beq
\kappa_H \, \leq \, \kappa_H^{max} \,\equiv \, \frac{c_\alpha^{2,max}}{\hat{c}_\alpha^{2,min}} \= \frac{1}{\hat{c}_\alpha^{2,min}} 
\eeq
If we use $1\sigma$ or $95\%$CL lower ranges of $\hat{c}_\alpha$ fit result, we obtain
\beq
\kappa_H^2 \= \frac{\Gamma^{tot}}{\Gamma^{tot}_{SM}} \, \leq \, \left\{ \bmat 1.36 \,(1\sigma) \\ 1.65\,(95\%CL). \emat \right.
\eeq
Likewise, we can solve the relation $\kappa_H = c_\alpha^2/ \hat{c}_\alpha^2$ for $BR_{nonSM}$ using \Eq{eq:nonsmwid}, and obtain
\beq
BR_{nonSM} \= 1 \- \frac{ \hat{c}_\alpha^4}{c_\alpha^4} ( 0.0857\kappa_g^2 + 0.9143 c_\alpha^2) \, \lesssim \, 1 \- \hat{c}_{\alpha,min}^4 \= \left\{ \bmat 0.27\,(1\sigma) \\  0.39\,(95\%CL). \emat \right.
\eeq

In obtaining these upper bounds, we have used built-in condition, $c_\alpha \leq 1$. If certain well-motivated theoretical assumptions on vector-boson couplings are made, one may obtain stronger bounds on non-standard decay width, e.g., see Refs.\cite{Dobrescu:2012td, Belanger:2013kya}.

Note that the singlet-like eigenstate can still have some non-standard decay modes if those modes are not be kinematically allowed for SM-like Higgs. These decays are not constrained from global fit of 125GeV data.

Let us consider a special case where $c_g = 0$, namely that the singlet scalar has interactions with new
charged particles that are colorless :  $\{ \, \alpha, \, c_\gamma, \, BR_{nonSM} \,\}$.  Then one has 
\beq
\kappa_\gamma^2 \= (c_\alpha - c_\gamma s_\alpha)^2, \quad \kappa_g^2 \= \kappa_{mix}^2 \= c_\alpha^2, \quad \kappa_H^2 \= \frac{c_\alpha^2}{1-BR_{nonSM}}
\eeq
giving the best-fit
\beq
\hat{c}_\alpha \= 0.990_{-0.0292}, \quad \hat{c}_\gamma \= -0.786^{+0.714}_{-0.637}, \quad \chi^2/\nu \= 11.1/8 \=1.39.
\eeq
Upper bound on non-standard branching ration is also obtained in the same way as above
\beq
BR_{nonSM} \, \leq \, 1- \hat{c}_{\alpha,min}^4 \= \left\{ \bmat 0.15 \,(1\sigma) \\ 0.24 \,(95\%CL). \emat \right. 
\eeq

\subsection{More dedicated models with multiple leptons or quarks} \label{sec:abc-fit}

Models of multiple vector-like leptons can couple both to Higgs and singlet scalar directly. This type of model has been studied to simultaneously enhance diphoton rate and tame resulting vacuum instability via singlet scalar threshold effects \cite{Batell:2012zw}. $b_\gamma$ is then modified in addition to $c_\gamma$. Thus one can consider following case :  $\{ \, \alpha,\, c_\gamma, \, b_\gamma \, \}$. 
However, it is clear that no unique solution of $b_\gamma, c_\gamma$ will be obtained because those two 
free parameters only enter into a single $\kappa_\gamma$. In terms of underlying model parameters, 
$b_\gamma$ and $c_\gamma$ are induced by different independent couplings. Thus, underlying model 
is also not well constrained from global fit. We do not study this modes furthermore in this paper.

\begin{table}[t] \centering \begin{tabular}{c||c|c}
\hline \hline
    Models        & Best-fit results  &  $\chi^2/\nu$\\
\hline \hline
SM         &  & $12.01/10 = 1.20$  \\
\hline \hline
universal modification &       &  \\  
 $(\hat{\kappa}_{univ}^2)$   & $(1.012)  $  & $11.96/9 =1.33$  \\
 $(BR_{nonSM})$ &  $\leq 18.8\%$ at 95\%CL \\
 $(\cos \alpha)$ & $\geq 0.904$ at 95\%CL\\
\hline
VL lepton, $W^\prime$, $S^\prime$ &   & \\
$(c_\alpha, c_\gamma)$ &   (0.98, -0.55) & $11.1/8=1.39 $   \\
\hline
VL quark &  &  \\
$(c_\alpha, c_g, c_\gamma)$ &  (0.947, -0.128, -0.313) & $11.1/7=1.58  $ \\
\hline
  &  &  \\
$(c_\alpha, c_\gamma, Br_{nonSM})$ &  $BR_{nonSM} \leq 24\%$ at 95\%CL & $11.1/8=1.39 $\\
\hline
  &  & $ $ \\
$(c_\alpha, c_g, c_\gamma, Br_{nonSM})$ &   $BR_{nonSM} \leq 39\%$ at 95\%CL & $11.1/7=1.58  $ \\
\hline \hline
singlet mixed-in $\hat{\kappa}$ &   & \\
$(\hat{\kappa}_g^2, \hat{\kappa}_\gamma^2, \hat{\kappa}_{mix}^2)$ & (1.03, 1.15, 0.942) &  $11.1/7=1.58 $ \\
\hline
singlet mixed-in theory  &  &  \\
$(\hat{c}_g, \hat{c}_\gamma, \hat{c}_\alpha)$ & (-0.176, -0.432, 0.971) & $ 11.1/7=1.58 $ \\
\hline \hline
\end{tabular} \caption{Summary of best-fit results with scalar mixing. If $BR_{nonSM}$ is included in fit, no unique solution is found, and its upper bound at 95\%CL is presented. Only central values of best-fit are shown, and errors can be found in text.}
\label{tab:c-fit} \end{table}

\section{Implications on the other scalar boson nearby 125 GeV Higgs boson} \label{sec:sing-near}

Interesting application of our global fit in terms of the coupling constants of interaction eigenstates is to 
derive further constraints on the extra scalar boson mixing with 125GeV resonance. We use the null results of SM Higgs searches at LEP and the $\gamma \gamma,\, ZZ$ resonance searches in the other mass range at LHC to illustrate the application. The evolution of loop-induced couplings will also be discussed. Recently reported diphoton bump at 136.5GeV observed by CMS will be briefly discussed at the end.

\subsection{Universal modification and LEP bounds}

 Universal modification scenario in terms of two parameters 
\beq  \{ \, \alpha ,\, BR_{nonSM}  \, \} \eeq
is the simplest scenario that can be constrained from LEP searches of light Higgs boson. $BR_{nonSM}$ 
can be relevant if $m_s \leq m_h/2$. Our discussion in this section, however, does not depend on whether $m_s \leq m_h/2$ or not because only $\hat{\kappa}_{univ}$ combination (not individual $\alpha$ or $BR_{nonSM}$) is constrained; see Sec.5.1 for discussion. 
LHC bounds will be discussed in the next subsection using another scenario although similar bounds 
can be derived in this case. As both production and decay relevant at LEP are proceeded by tree-level induced coupling. Thus, we do not need to discuss the mass dependences of couplings here -- we refer to next subsection for this discussion. With these, the signal strengths of $H_2$ is universally modified as that of $H_1$.

\begin{figure} \centering
\includegraphics[width=\textwidth]{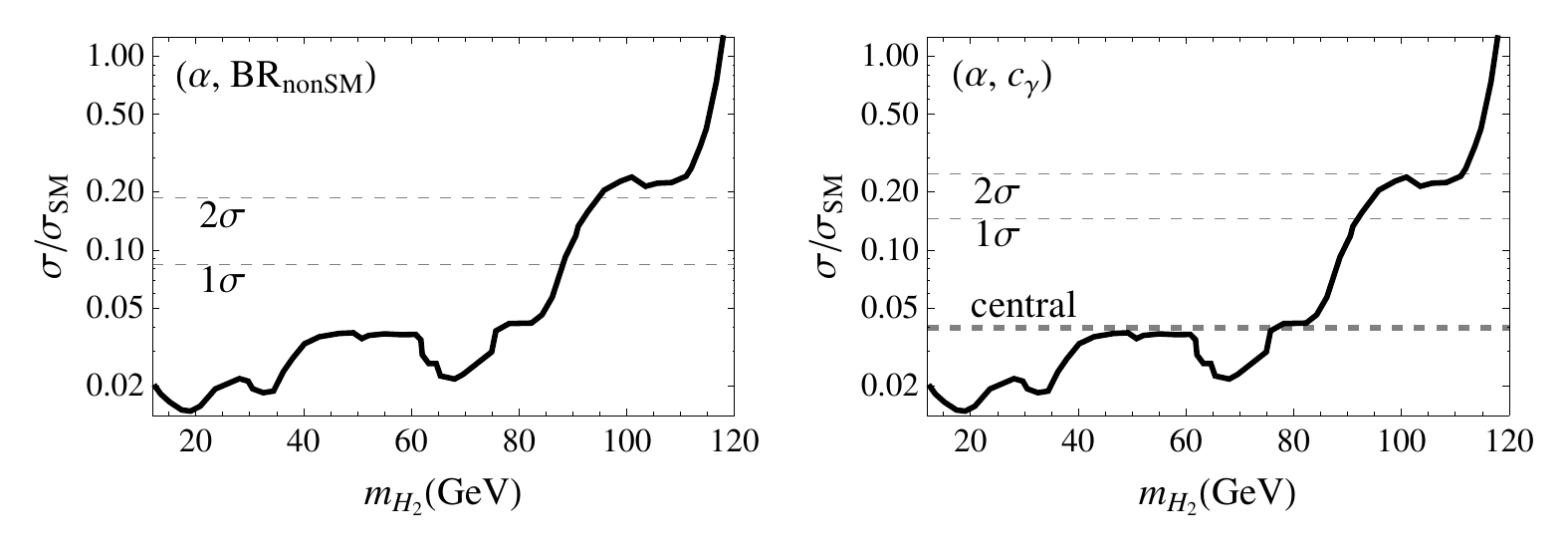}
\caption{95\%CL LEP upper bound on the signal strength 
$e^+ e^- \to Z^{(*)} \to Z^{(*)}h (\to b\bar{b}, \tau\tau)$ \cite{Barate:2003sz} is shown as thick line. 
Regions below dashed lines are preferred by global fit at given confidence level. (Left): global fit in terms of $\alpha$ and/or $BR_{nonSM}$. (Right): global fit in terms of $\alpha$ and $c_\gamma$. $c_\gamma$ is fixed to its best-fit value in the plot.}
\label{fig:lep-light}
\end{figure}

LEP1 looked for Higgs via Bjorken process $Z \to Z^* h$ followed by $Z \to \ell^+ \ell^-, \nu \bar{\nu}, b\bar{b}, \tau\tau$ and $h \to b\bar{b}, \tau \tau$. LEP2 looked for Higgs via $e^+ e^- \to Zh$ \cite{Barate:2003sz}. Signal strengths in all cases are globally parameterized by
\beq
\frac{\sigma}{\sigma_{SM}} \= \hat{\kappa}_{univ}^4 \= c_\alpha^2 ( 1-BR_{nonSM}).
\eeq
The LEP bounds on the signal strength can be interpreted as bounds on the parameter $\hat{\kappa}_{univ}$. The $1,2 \sigma$ favored region of global fit is overlayed with LEP bounds in \Fig{fig:lep-light} left panel. Since SM is consistent with best-fit results, no strong bound can be derived yet. However, best-fit results can already constrain heavy mass regime better than LEP.

\subsection{Scalars enhancing the diphoton rate with $c_\gamma$}

This subsection discusses a useful application of our global fits and Lagranigan parameterization introduced in this paper. We will evolve the coupling constants $b_i$ and $c_i$ fitted at 125GeV to the mass of the singlet $m_{H_2}\ne 125\GeV$ in order to extract physical couplings of $H_2$ bosons, and discuss if these couplings can be further constrained from other $\gamma \gamma$\cite{ATLAS:diphoton-res}, $ZZ$\cite{ATLAS:data2} resonance searches at collider. To this end, we will especially have to discuss how mass dependences of loop-induced couplings can be evolved properly. All other couplings are assumed to be tree-level generated and to be constant over the mass range. We explicitly write the mass dependences of loop-induced couplings: $c_{g,\gamma}(m)$ and  $b_{g,\gamma}(m)$. Recall that these parameters are defined and determiend at $m=m_{H_1}=125$GeV by global fits. 

\begin{figure} \centering
\includegraphics[width=\textwidth]{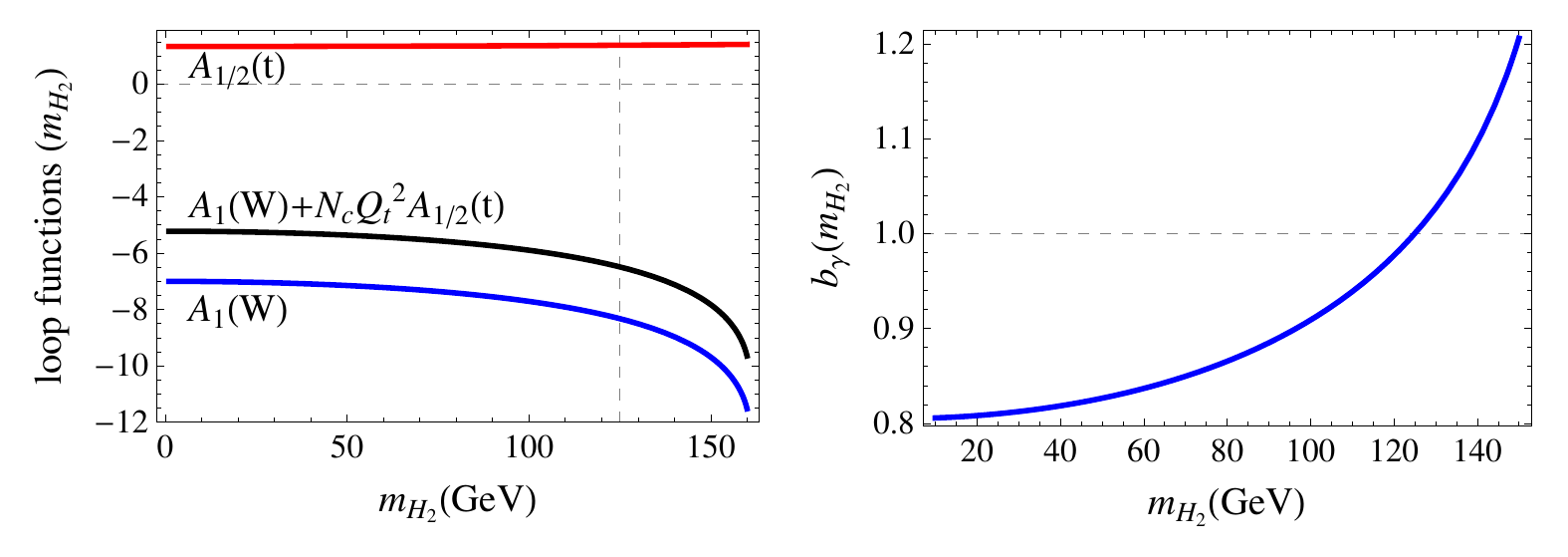}
\caption{(Left): Loop functions of $W$ boson, $A_1(W)$, and top quark, $A_{1/2}(t)$, are evaluated at mass scale $m_{H_2}$. SM Higgs coupling to photons corresponds to black line while coupling to gluon is proportional to $A_{1/2}(t)$. (Right): $b_\gamma(m_{H_2})$ relevant to the coupling of $H_2$ boson to photons as defined in \Eq{eq:s-photon}.}
\label{fig:loop-func}
\end{figure}

Focusing on $m_{H_2} \lesssim 160\GeV$ allows us to meaningfully extract loop-induced couplings of $H_2$. This is technically useful because the mass dependences of $c_{g,\gamma}$ and $b_g$ coming from the loops of heavy particles -- top quarks and extra new vector-like leptons or quarks -- are small. See \Fig{fig:loop-func} that top quark's loop function does not change much in this mass range; thus, so do heavier particles' loop functions. This is very useful simplification as $c_g$ and $b_g$ enter via $gg$ fusion production of $H_2$, and thus $gg$ fusion is independent on mass scale -- it is simply rescaled by mixing angle regardless of $m_{H_2}$. New particles, of course, could be lighter than $H_2$ or $2m_W$, but we do not consider such a case given that collider bounds generally predict heavier charged (and/or colored) particles. On the other hand, $b_\gamma(m_{H_2})$ scales mildly with $m_{H_2}$ in this mass range due to the contribution from lighter $W$ boson as shown in \Fig{fig:loop-func}. The $W$ loop function could have behaved more rapidly and obtained imaginary part if $H_2$ were heavier than about $2m_W$. Also, $H_2$ could have dominantly decayed to $WW^*$ making diphoton modes irrelevant. In all, we assume that $c_{g,\gamma}, b_g$ are mass-independent, whereas $b_\gamma$ is properly evolved from 125GeV to $m_{H_2}$ as will be explained in \Eq{eq:s-photon}. For simplicity, we assume that $H_2$ does not have extra non-standard decay modes.

We consider a two-parameter fit
\beq
\{ \, \alpha, \, c_\gamma \, \}.
\eeq
Decay widths of 125GeV Higgs boson is denoted by $\kappa_i^2$ as usual
\beq
\kappa_\gamma^2 \= \left. \frac{ \Gamma(H_1 \to \gamma \gamma) }{ \Gamma(h \to \gamma \gamma)_{SM} } \right|_{125\GeV} \= ( b_\gamma(125) c_\alpha \- c_\gamma(125) s_\alpha )^2 \= (c_\alpha \- c_\gamma s_\alpha)^2.
\eeq
where $b_\gamma(125)=1$ and $c_\gamma(m)=c_\gamma$ constant as discussed. The $H_2$ decay width with respect to SM width at $m_{H_2}$ is written with some care by
\bea
\zeta_\gamma^2 \= \left. \frac{ \Gamma(H_2 \to \gamma \gamma) }{ \Gamma(h \to \gamma \gamma)_{SM} } \right|_{m_{H_2}} &= &  \frac{ \Gamma(H_2 \to \gamma \gamma) |_{m_{H_2}} }{ \Gamma(h \to \gamma \gamma)_{SM} |_{125}} \, \cdot \, \frac{ \Gamma(h \to \gamma \gamma)_{SM} |_{125} }{ \Gamma(h \to \gamma \gamma)_{SM} |_{m_{H_2}}} \nonumber \\
&=& \left( c_\gamma(m_{H_2}) c_\alpha \+ b_\gamma(m_{H_2}) s_\alpha \right)^2 \cdot \left( \frac{ b_\gamma(125) }{ b_\gamma(m_{H_2}) } \right)^2  \\
&=&  \left( \frac{c_\gamma(m_{H_2})}{b_\gamma(m_{H_2})} c_\alpha \+  s_\alpha \right)^2 
\label{eq:sin-diph-width}\eea
where in the second line, we used the fact that $c_i$ and $b_i$ are defined with respect to corresponding 125GeV SM Higgs couplings. $c_\gamma(m_{H_2}) = c_\gamma(m_{H_1}) = c_\gamma$ constant. $b_\gamma(m_{H_2})$ is given by
\bea
b_\gamma(m_{H_2})^ 2 &=& \left| \frac{1}{A_{SM}^\gamma} \left( A_1 ( m_{H_2} /4m_W^2 ) \+ N_c Q_t^2 A_{1/2}(m_{H_2}/4m_t^2) \right) \right|^2  \nonumber\\
&=& \left| \frac{  A_1 ( m_{H_2} /4m_W^2 ) \+ N_c Q_t^2 A_{1/2}(m_{H_2}/4m_t^2) }{  A_1 ( m_{H_1} /4m_W^2 ) \+ N_c Q_t^2 A_{1/2}(m_{H_1}/4m_t^2) } \right|^2
\label{eq:s-photon} \eea
and is plotted in the right panel of \Fig{fig:loop-func}. Decay widths of other decay modes are simply rescaled by mixing angle
\beq
\zeta_i^2 \= \frac{ \Gamma(H_2 \to i) }{ \Gamma(h \to i)_{SM} } \= s_\alpha^2.
\eeq
Total decay width of $H_2$ is then also approximately 
\beq
\zeta_H^2 \= \frac{\Gamma(H_2)^{tot}}{\Gamma(h)^{tot}_{SM}} \, \simeq \, s_\alpha^2.
\eeq

\begin{figure} \centering
\includegraphics[width=0.95\textwidth]{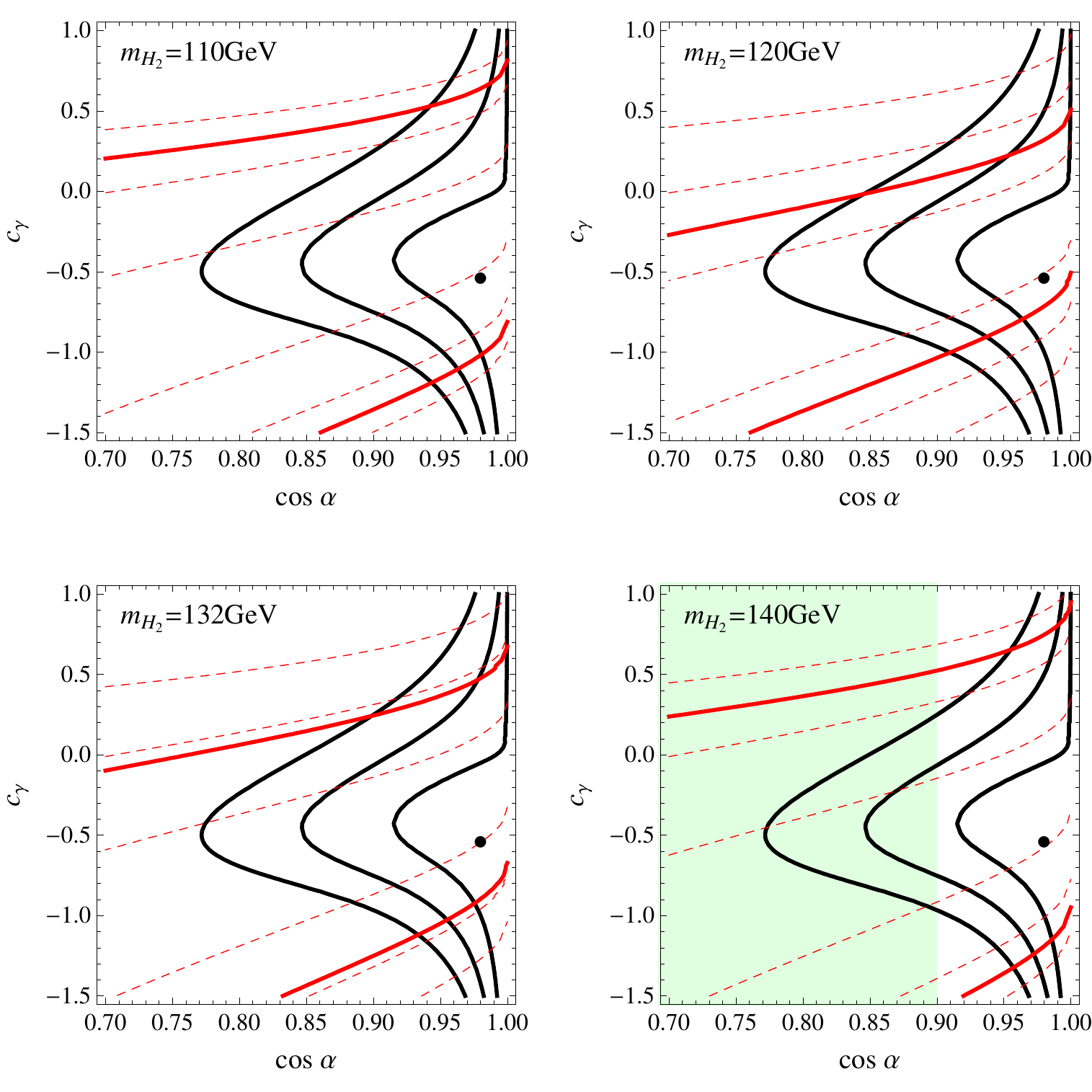}
\caption{In addition to $1,2,3 \,\sigma$ favored region of global fit of $\{ \alpha, c_\gamma \}$ as black lines, contours of $\zeta_\gamma^2 = R(\sigma(pp\to H_2) \times BR(H_2 \to \gamma \gamma))$ are shown as red-dashed. From two central lines, $\zeta_\gamma^2 = 0.1, \, 0.5, \, 1$. Regions between two thick red lines are allowed from 95\%CL bound of diphoton resonance search at LHC \cite{ATLAS:diphoton-res}. Shaded region in the last plot is excluded from $ZZ^* \to 4 \ell$ resonance search at LHC by 95\%CL \cite{ATLAS:data2}. This bound is too weak to be shown in other panels.}
\label{fig:diphoton-bound}
\end{figure}

Similarly to previous case, LEP bound overlapped with favored region of global fit results in \Eq{eq:fit-alcgam} is shown in the right panel of \Fig{fig:lep-light}. If future data would predict the same central value with smaller error, we might obtain a band of favored region which can narrow down the allowed mass range. Diphoton resonance search \cite{ATLAS:diphoton-res} and $ZZ^{(*)} \to 4\ell$ resonance search at LHC \cite{ATLAS:data2} constrain following signal strength parametrizations
\beq
R( \, \sigma(pp \to H_2) \times BR(H_2 \to \gamma \gamma) \,) \= \hat{\zeta}_g^2 \hat{\zeta}_\gamma^2 \= \zeta_\gamma^2
\ceq
R( \, \sigma(pp \to H_2) \times BR(H_2 \to ZZ^*) \,) \= \hat{\zeta}_g^2 \hat{\zeta}_Z^2 \= s_\alpha^2
\label{eq:sin-diph-signal}\eeq
where $\hat{\zeta}_i^2 = \zeta_i^2 / \zeta_H$ as before. Bounds on the parameter space from these resonance searches and global fits are shown in \Fig{fig:diphoton-bound}. As search bounds vary with the mass of a resonance, we pick four representative results in the plot. Contours of constant $\zeta_\gamma^2$ are also shown for reference. SM is consistent with both global fit and resonance searches. Nonetheless, some region preferred by global fits is excluded by resonance searches. 

\subsection{Addendum with 136.5GeV diphoton accumulation}

We also carry out a quick and interesting aside. CMS has recently observed the slight excess of diphoton resonance data at around 136.5GeV~\cite{CMS:136}. ATLAS has not so far observed corresponding excess. No detailed CMS data is available yet. But we assume following data $\mu_{\gamma \gamma}(136.5\GeV) = 2.3 \pm 0.45$ and illustrate how our method can be applied to interpret this data. 

We use $\{ c_\alpha, \, c_\gamma \}$ ansatz to fit both 125GeV and 136.5GeV CMS data simultaneously. ATLAS data is not included. From previous subsection, 136.5GeV diphoton signal strength (compared to that of 136.5GeV SM Higgs rate) is given by \Eq{eq:sin-diph-width} and \Eq{eq:sin-diph-signal}. Here, $b_\gamma(136.5\GeV) = 1.069$ and $c_\gamma(136.5\GeV)=c_\gamma(125\GeV)=c_\gamma$ is constant. Using these, our best fit yields
\beq
c_\gamma \= 1.54, \quad c_\alpha \= 0.997, \qquad \chi^2/\nu \= 1.65/5 \= 0.33
\eeq
which can be compared with best-fit results without 136.5GeV data: $c_\gamma \= 0.173, c_\alpha = 0.942, \chi^2/\nu \= 1.13/4 \= 0.28$ (only CMS) and $c_\alpha \= 0.98, c_\gamma \= -0.55, \chi^2/\nu \= 11.1/8 \=1.39$ (both CMS and ATLAS as reported in \Eq{eq:fit-alcgam}). Notably, sizable $c_\gamma$ is preferred compared to CMS-alone results while positive $c_\gamma$ is preferred as opposed to CMS+ATLAS results. The latter may imply that ATLAS 125GeV diphoton enhancement and CMS 136.5GeV diphoton enhancement may not consistently coexist in the model of singlet extension (see Sec.\ref{sec:bi-fit} for disucssions on general trends of ATLAS and CMS datasets). We comment that we restrict $s_\alpha>0$ throughout this paper which doesn't matter if only 125GeV data is used in the global fit, but changing the sign of $s_\alpha$ yields somewhat worse fit here.

\section{Conclusions} \label{sec:conc}

In this paper, we performed comprehensive analysis of the current LHC data on the 125 GeV 
Higgs-like resonance assuming that the SM Higgs properties can be modified either by new physics 
in the loop or by a mixing with a nearby singlet scalar boson.  Our approach is conceptually more 
general than other approaches based on the effective Lagrangian with higher dimensional operators 
with the SM Higgs boson $h$.  We imposed the SM gauge symmetry 
$SU(3)_c \times SU(2)_L \times U(1)_Y$ instead of $SU(3)_c \times U(1)_{\rm em}$, and distinguish the nature of the SM Higgs $h$ and a singlet scalar $s$. In practice, our fitting procedure is identical  to others, but the interpretation of the results could be different.
We could obtain more detailed informations on possible new physics impacts on the SM Higgs properties 
within our approach. 

We presented a number of interesting class of BSM's where  new singlet scalar bosons appear with 
couplings to the SM fields as well as to some new fields such as new charged vector bosons, 
vectorlike fermions or hidden sector dark matters, etc.. The singlet scalar boson(s) mix with
the SM Higgs boson, and thus modify the Higgs properties. 

To parametrize these general modification due to a single singlet scalar, we introduce separate coupling constants to the Higgs and singlet interaction eigenstates. Their mixing is parameterized by a single mixing anlge. We later have also discussed how this paraemterization can be further used to constrain the singlet scalar interactions from other resonance searches at collider.

We carried out global fits assuming various ansatzs representing BSM's in literature. Interestingly enough, the LHC data after Moriond 2013, according to our global fit, implies that the SM  gives the best $\chi^2 / \nu = 1.20$ 
when both ATLAS and CMS data are used for the fit. All other models we considered, such as models with 
extra vector-like fermions or charged vector bosons that were considered for the enhancement of 
$H \rightarrow \gamma\gamma$ or new sequential fermions or 2 HDM's, or models with extra singlet scalar
bosons, yield larger $\chi^2/\nu$ compared with the SM (see Tables~\ref{tab:fit-bi} and \ref{tab:c-fit}).  
If we consider the ATLAS and the CMS  data separately and perform the $\chi^2$ minimization fit,
the CMS data seems to prefer a singlet scalar boson with some suppression of the signal 
strengths. If the SM is assumed, $\chi^2/\nu = 0.466$ is obtained for the CMS-only which is smaller
than ATLAS + CMS SM fit ($\chi^2 / \nu = 1.20$).  On the other hand, the ATLAS data can be better fitted with BSM in terms of 
$b_i \neq 1$, but if the SM is assumed, ATLAS's $\chi^2 / \nu = 1.94$ becomes worse than the ATLAS + CMS SM fit.
Of course, the number of degrees of freedom in our fits maybe too small
to make a strong conclusion at the moment, but this observation may be suggestive. 

In order to make a stronger conclusion, we have to await more data accumulation from the next run of the 
LHC at 14 TeV, with better statistics and systematics, as well as better determination of the signal 
strengths for different Higgs production channels.  In many BSM's we presented in this paper, 
there appears a second neutral Higgs-like scalar boson, a mixture of the SM Higgs boson $h$ and 
a singlet scalar $s$.   It would not be easy to search for this second scalar, because it should be mostly 
a singlet scalar, considering  the current data on the 125 GeV Higgs-like resonance.  
Still, search for this second scalar is clearly warranted  because it may be the only way to probe 
dark sector made of the SM singlet fields at colliders.

\acknowledgments
The authors would like to thank Dongwon Jung, Kyoungchul Kong, Hyunmin Lee 
and Eibun Senaha for their useful comments. SJ thanks KIAS Center fro Advanced Computation for providing computing resources.
This work was supported by NRF Research Grant 
2012R1A2A1A01006053 (PK), 2011-0016554 (SC) and by 
National Research Foundation of Korea(NRF) grant funded by the Korea 
government(MSIP) No. 2009-0083526 through  Korea Neutrino Research Center 
at Seoul National University (PK). This paper is registered as a prinprint KIAS-P13038.



\end{document}